\documentclass[aps,prd,showpacs,twocolumn,superscriptaddress,nofootinbib,preprintnumbers]{revtex4-1} 

\makeatletter
\def\p@subsection{}
\makeatother

\usepackage{color,graphicx,amsmath,amssymb,lipsum}
\usepackage[colorlinks=true,citecolor=blue,linkcolor=blue,urlcolor=blue, backref=false,pdfborder={0 0 0}]{hyperref}

\usepackage[normalem]{ulem}
\usepackage[utf8]{inputenc} 

\usepackage{float}

\usepackage{ulem}
\usepackage{diagbox}

\usepackage[dvipsnames]{xcolor}

\newcommand{\be}{\begin{equation}}
\newcommand{\ee}{\end{equation}}
\newcommand{\beqa}{\begin{eqnarray}}
\newcommand{\eeqa}{\end{eqnarray}}

\renewcommand\t{\theta}

\renewcommand\L{\Lambda}

\def\d{\partial}
\newcommand{\bseq}{\begin{subequations}}
\newcommand{\eseq}{\end{subequations}}

\renewcommand{\ln}{\mathop{\rm ln}\nolimits}

\def\gsim{\raise0.3ex\hbox{$\;>$\kern-0.75em\raise-1.1ex\hbox{$\sim\;$}}}
\def\lsim{\raise0.3ex\hbox{$\;<$\kern-0.75em\raise-1.1ex\hbox{$\sim\;$}}}

\def\beqn#1{\begin{equation}\label{#1}}
\def\eeqn{\end{equation}}

\def\beqa#1{\begin{eqnarray}\label{#1}}
\def\eeqa{\end{eqnarray}}

\def\Z2{$\mathcal{Z_2}$}

\newcommand {\ignore}[1]{}

\begin{document}

\preprint{INR-TH-2020-025}
\preprint{TTK-20-11}

\title{$H_0$ tension or $T_0$ tension?}

\author{Mikhail M. Ivanov}\email{mi1271@nyu.edu}\affiliation{Center for Cosmology and Particle Physics, Department of Physics, New York University, New York, NY 10003, USA}
\affiliation{Institute for Nuclear Research of the
Russian Academy of Sciences, \\ 
60th October Anniversary Prospect, 7a, 117312
Moscow, Russia
}
\author{Yacine Ali-Ha\"{ı}moud}\email{yah2@nyu.edu}\affiliation{Center for Cosmology and Particle Physics, Department of Physics, New York University, New York, NY 10003, USA}
\author{Julien Lesgourgues}\email{julien.lesgourgues@physik.rwth-aachen.de}\affiliation{Institute for Theoretical Particle Physics and Cosmology (TTK)\\\ RWTH Aachen University, D-52056 Aachen, Germany
}

\begin{abstract} 
We study if the discrepancy between the local and cosmological measurements of the Hubble constant $H_0$ can be reformulated as a tension in the cosmic microwave background (CMB) monopole temperature $T_0$. The latter is customarily fixed to the FIRAS best-fit value in CMB data analyses. Although this value was confirmed by several independent experiments, it is interesting to see how much parameter constraints depend on this prior. We first provide a detailed pedagogical description of the $T_0$ effects on cosmological observables. We show that the recombination history and transfer functions do not depend on $T_0$, provided they are parametrized by the energy scale rather than redshift, and at fixed dark matter and baryon densities per CMB photon. 
Thus, $T_0$ is only a property of the observer, quantifying the amount of expansion between key cosmological events and today. As a consequence, the sole effect of $T_0$ on small-scale primary CMB anisotropies is through the angular diameter distance to the epoch of last scattering, resulting in a near-perfect degeneracy between $T_0$ and $H_0$. This geometric degeneracy is partially lifted by the late-time integrated Sachs-Wolfe effect and CMB lensing.
Still, Planck data alone is consistent with a broad region 
in the $H_0-T_0$ plane, implying
that removing the FIRAS prior on $T_0$ can make Planck and SH0ES less discrepant, without introducing new physics beyond $\Lambda$CDM. One may break the degeneracy by combining Planck with SH0ES, yielding an independent measurement of $T_0$, which happens to be in $3\sigma$ tension with FIRAS. Therefore, the Hubble tension indeed can be recast into the $T_0$ tension. The agreement with FIRAS is restored when combining Planck with the baryon acoustic oscillation data instead of SH0ES. Thus, the tension between SH0ES and cosmological measurements of $H_0$ persists even if we discard the FIRAS $T_0$ prior.
\end{abstract}

\maketitle

\section{Introduction and Summary}

The disagreement between the value of the Hubble constant $H_0$ measured by different methods 
(the so-called ``Hubble tension'')
has recently become a hot topic in cosmology. 
On the one hand, local measurements using the Cepheid-calibrated supernovae~\cite{Riess:2019cxk,Reid:2019tiq}
and strong lensing time-delays~\cite{Wong:2019kwg} yield a number
around $74$~km/s/Mpc.
On the other hand, the Planck cosmic microwave background 
radiation (CMB) data \cite{Aghanim:2018eyx}, various large-scale structure (LSS) probes \cite{Abbott:2017smn,Cuceu:2019for,Schoneberg:2019wmt,Ivanov:2019pdj,Leonardo:2019me,Colas:2019ret,Philcox:2020vvt}, as well as the local measurements based on 
the inverse distance ladder technique \cite{Aubourg:2014yra,Lemos:2018smw}
favour independently of each other a value close to $68$~km/s/Mpc.
The Hubble tension might be the result of unaccounted systematics\footnote{See, e.g. Ref.~\cite{Rigault:2014kaa} in the context of 
the Cepheid-calibrated supernovae and Refs.~\cite{Kochanek:2019ruu,Blum:2020mgu}
for discussions regarding the strong lensing time delays.
} 
or a manifestation of new exotic physics (see Ref.~\cite{Knox:2019rjx} for a review).
Therefore, it is imperative to scrutinize various choices made
in the analysis of each dataset.
One of such assumptions is the value of the CMB monopole temperature $T_0$, which
is typically fixed when fitting the Planck CMB likelihoods.
The rationale behind this choice is that $T_0$ has been measured from a combination of the COBE/FIRAS
data, molecular lines, and balloon-borne experiments
with an outstanding precision~\cite{Fixsen:1996nj,2009ApJ...707..916F},
\be 
\label{eq:firas}
T_{0,{\rm FIRAS}}=(2.72548\pm 0.00057)~\text{K}\,.
\ee
For simplicity, we will call this measurement just ``FIRAS''. Note that before FIRAS, $T_0$ had been measured already with good accuracy by COBRA \cite{PhysRevLett.65.537}, and that there exist other independent measurements of $T_0$ from the observation of carbon monoxide absorption lines in quasar spectra \cite{2011A&A...526L...7N}, all compatible with FIRAS.\footnote{It is worth mentioning that there are further prospects to re-measure $T_0$ with experiments aimed at CMB spectral 
distortions. In particular, PIXIE \cite{Kogut:2011xw} would allow 
one to reduce the uncertainty of the $T_0$
measurement down to $\sigma(T_0)\sim 10^{-7}$~K \cite{Abitbol:2017vwa}, which is 3 orders of magnitude better than FIRAS. 
}

The effect of $T_0$ on CMB anisotropies was studied in Refs.~\cite{Chluba:2007zz,Hamann:2007sk,Yoo:2019dyl,Ade:2015xua,DiValentino:2016foa}. These past studies have either focused on the impact of uncertainties in $T_0$ on cosmological parameters inferred from CMB anisotropy data \cite{Chluba:2007zz, Hamann:2007sk, Yoo:2019dyl}, on combining current CMB anisotropy data with external datasets to measure $T_0$ \cite{Ade:2015xua}, or on determining whether future CMB-anisotropy experiments might be able to measure $T_0$ \cite{DiValentino:2016foa}. In this paper, we show, for the first time, that $T_0$ can be measured from \emph{current} CMB anisotropy data \emph{alone}. We moreover study whether removing the FIRAS prior on $T_0$ can alleviate the Hubble tension.
Since the monopole temperature can be seen as a proxy for the age of the Universe, just like $H_0$, the two quantities should be nearly perfectly degenerate. 
Clearly, the CMB measurement of $H_0$ must be influenced by the $T_0$ prior.
Naively, this offers a tempting way to resolve the Hubble tension without new physics
and any additional parameters beyond those already contained in the base $\L$CDM model.

Aiming at restoring the agreement between Planck and SH0ES,
we have reanalyzed the final Planck 2018
data without fixing the CMB monopole temperature.
As expected, we have found that the CMB data exhibit a clear $H_0-T_0$ degeneracy. 
However, even in this extended model, the SH0ES measurement remains in substantial ($\sim 3\sigma$) tension with Planck.
Thus, combining Planck and SH0ES gives an \textit{independent measurement} of $T_0$. 
This value happens to be in $\sim 3\sigma$ tension with the FIRAS measurement. However, one can break the geometric CMB degeneracy between $T_0$ and $H_0$
equally well with
low-redshift baryon acoustic oscillation (BAO) data. 
This leads to a different measurement of $T_0$
that agrees with FIRAS, while being in $\sim 3\sigma$ tension with SH0ES.

Thus, interestingly, the Hubble tension between Planck+FIRAS and SH0ES can be fully reformulated as a $T_0$ tension between Planck+SH0ES and FIRAS. However, when BAO data is taken into account, the very good agreement between the measurements of 
Planck, FIRAS and BAO suggests that, as long as the $\L$CDM model is assumed, SH0ES is the outlier.\footnote{Note that, in principle, $H_0$ can also be measured from $T_0$ 
in a model-independent way
via the observation of the CMB
monopole cooling~\cite{Abitbol:2019ewx}.}

Another goal of this paper is 
to clarify the physical effect of $T_0$ on cosmological observables.
In past works \cite{Chluba:2007zz,Hamann:2007sk}, $T_0$ was varied while 
keeping $H_0$ and the
\textit{current} energy densities of the baryons and dark matter $\omega_b$ and $\omega_{cdm}$ fixed. 
However, recombination physics and the early-time CMB
anisotropies are fully determined 
by the baryon-to-photon and CDM-particles-to-photon ratios, 
which results in almost perfect 
degeneracies  
between $H_0,\omega_b,\omega_{cdm}$ and $T_0$ \cite{DiValentino:2016foa}. Thus, varying $T_0$ with fixed 
$H_0,\omega_{cdm}$ and $\omega_b$ 
is not very informative. Instead, we show that it is more physically meaningful to study the effect of $T_0$ while keeping the ratios $\omega_{cdm}/T_0^3$ and $\omega_b/T_0^3$ constant. With these parameters fixed, the \emph{temperature} of recombination is fixed, and a change of $T_0$ \emph{mostly} amounts to changing the angular diameter distance to the last scattering surface, thus the angle $\theta_s$ that it substends, resulting in a strong degeneracy with $H_0$. In addition, however, a change in $T_0$ at fixed $\theta_s$ changes the time elapsed since matter-$\Lambda$ equality, thus the late integrated Sachs-Wolfe effect and gravitational lensing. Through these late-time effects, we show that one can extract a $4\%$ measurement of $T_0$ from Planck data \emph{alone}, contrasting with the standard lore that this could not be done without external datasets
\cite{Ade:2015xua, DiValentino:2016foa}.

The remainder of this paper is structured as follows. 
We start with the discussion of our datasets in Sec.~\ref{sec:data}.
We give some theoretical background in Sec.~\ref{sec:met}, whereas Sec.~\ref{sec:res} contains our main results. 
Finally, we
draw conclusions in Sec.~\ref{sec:concl}.

\section{Data}
\label{sec:data}

For CMB anisotropy data, we use the Planck baseline TTTEEE~+~lowE~+~lensing likelihood\footnote{We stick here to the naming conventions of the Planck collaboration, in which ``TTTEEE~+~lowE'' stands for high-$\ell$ TT, TE, EE data combined with low-$\ell$ TT, EE data; while lensing refers to the measurement of the lensing deflection spectrum based on 4-point correlation functions of CMB maps \cite{Aghanim:2018eyx}.} from the 2018 data release \cite{Aghanim:2018eyx} as implemented in \texttt{Montepython v3.0} \cite{Brinckmann:2018cvx}, see Ref.~\cite{Aghanim:2019ame}
for likelihood details. 
Since the standard recombination code \texttt{recfast} \cite{Seager:1999bc} uses fudge functions calibrated for a fixed $T_0$,
we use the more flexible and accurate code \texttt{HyRec} \cite{2010PhRvD..82f3521A, 2011PhRvD..83d3513A} that does not rely on any fiducial cosmology.
In addition to the cosmological parameters, we vary 21 Planck
nuisance parameters that capture various instrumental and systematic effects \cite{Aghanim:2019ame}.

As for the SH0ES data, we will use a Gaussian prior on $H_0$ 
derived from the most recent measurements by the SH0ES collaboration \cite{Reid:2019tiq},
\be
H_0=73.5 \pm 1.4\quad {\rm km/s/Mpc}\,.
\ee
Moreover, we will employ BAO data from the 
BOSS data release 12 \cite{Alam:2016hwk}. 
In principle, one can derive better constraints from the full
BOSS likelihood that includes the shape information as well \cite{Chudaykin:2020aoj}.
However, the compressed likelihood  involving only the 
BAO scale will be sufficient 
if our goal is to break geometric degeneracies 
(see Ref.~\cite{Ivanov:2019hqk} for a discussion on the role of the BOSS data in combination with Planck).
In principle, one can also use more BAO 
measurements, e.g. from the~Ly-$\alpha$ \cite{Bourboux:2017cbm} and quasar data~\cite{Ata:2017dya}. 
However, 
the single most constraining BAO dataset 
from BOSS DR12
will be enough for the purposes of our paper.

The CMB monopole temperature $T_0$ has already been constrained independently of FIRAS in the Planck 2015 analysis \cite{Ade:2015xua}, which gave 
the following result from the combination of TT, TE, EE and BAO data,
\be
T_0 = \left(2.718 \pm 0.021\right)~{\rm K}\,. 
\ee
In this paper, it was already pointed out that the Planck data have a strong 
geometric degeneracy between $H_0$ and $T_0$, which can be broken by BAO measurements.
In the next sections we will 
explain in detail the origin 
of this degeneracy, and show how it can also be broken by SH0ES.

\section{Theory Background}
\label{sec:met}

\subsection{Cosmological model and parameters}

In this paper, we use geometric units $G = c = 1$. For short, we define the constant rate
\be
\gamma_{100} \equiv \frac{H_0}{h} = ~ 100 ~ \textrm{km/s/Mpc}.
\ee
In what follows we will focus on the Planck baseline $\L$CDM model \cite{Aghanim:2018eyx}. Specifically, we assume a spatially flat Universe with a cosmological constant $\L = 8 \pi \rho_\L$, containing thermal photons at temperature $T_\gamma$, cold dark matter, non-relativistic baryons, two massless neutrinos and a single massive neutrino with minimal mass $m_\nu = 0.06$ eV (but our entire discussion would hold in the more realistic case of three non-zero masses). We assume that neutrinos have the standard temperature $T_\nu = (4/11)^{1/3} T_\gamma$, and that the small non-thermal distortions to their spectrum can be accounted for with an effective number of relativistic degrees of freedom $N_{\rm eff} = 3.046$. 
 
Moreover, we assume scalar adiabatic initial conditions characterized by a simple power-law power spectrum. 
The cosmological model is then entirely determined by 7 parameters:

$\bullet$ 2 parameters determining the initial conditions: the amplitude $A_s$ 
at a reference scale $k_{\rm P}=0.05~\text{Mpc}^{-1}$ and tilt $n_s$ of the spectrum of primordial curvature fluctuations ${\cal R}$, $\Delta^2_{\cal R} = A_s (k/k_{\rm P})^{n_s -1}$.

$\bullet$ 4 independent parameters determining the matter and energy content: the cosmological constant $\L$, the present-time radiation temperature $T_0$ and the present-time baryon and cold dark matter densities $\rho_{b,0}, \rho_{c, 0}$, or their dimensionless versions $\omega_i \equiv \frac{8 \pi}{3} \rho_{i, 0}/ \gamma_{100}^2$. Instead of $\L$, one can equivalently use the Hubble parameter $H_0$ or the angular scale $\t_s$ of the sound horizon at last scattering. 
This phenomenological parameter 
is designed to approximate the
observed angular scale of the CMB acoustic peaks.\footnote{By  definition \mbox{$\theta_s = r_s/((1+z_{*})D_A(z_{*}))$} ($r_s$ is the comoving sound horizon at 
the redshift of recombination  $z_*$, $D_A(z_*)$ is the angular diameter distance to the last scattering surface). 
It should be borne in mind that this definition 
is somewhat ambiguous because recombination is not an instantaneous process.} 

$\bullet$ 1 astrophysical parameter: the optical depth  $\tau_{\rm reio}$ to reionization. 

\subsection{How can we know whether $T_0$ is measurable?}\label{sec:is_measurable}

It is standard in CMB anisotropy analyses to set $T_0$ to the mean value measured by FIRAS. Here we shall instead take $T_0$ as a free parameter. To check whether $T_0$ is actually measurable, we want to understand whether the impact of a variation of $T_0$ on cosmological observables can be absorbed by a rescaling of other parameters. Such observables depend on the evolution of several quantities that can be expressed as a function of different measures of time: proper time $t$, conformal time $\eta$, redshift $z$, the scale factor $a=1/(1+z)$, the energy scale, etc. We want to study whether we can ``vary $T_0$ and other parameters while keeping the cosmological evolution unchanged'', but given the previous remark, this could have several different meanings. The most relevant options are: \\

$(i)$ To maintain a fixed expansion history \emph{relative to today}. This choice may also be sensible, because several observables depend on characteristic scales measured relatively to lengths today, and on the amount of expansion between characteristic times and today. For instance, the amplitude of matter perturbations at some redshift $z$ relative to their amplitude today depends on the value of their wavelength relative to the Hubble radius today; the angle under which we see features in correlation functions depends on their size relative to the angular diameter distance, that we compute by integrating the expansion history relatively to the present time; the late integrated Sachs-Wolfe effect depends on the amount of expansion between matter-to-$\Lambda$ equality and today; etc. To explore such a degeneracy, we should try to express all relevant quantities as a function of the redshift $z$ (i.e.~of the scale factor relative to today), and to show that if cosmological parameters are scaled properly when $T_0$ is varied, the quantities over which cosmological observables depend remain invariant.  This could be achieved to some extent  by fixing the parameter combinations $\omega_b / T_0^4$, $\omega_{cdm} / T_0^4$, $\Lambda / T_0^4$, $N_\mathrm{eff}$, etc. \\

$(ii)$ To maintain cosmological evolution (of background, thermodynamics and perturbed quantities) as a function of \emph{absolute energy scales}. This choice is motivated by the fact that important phenomena like nucleosynthesis or recombination are determined by absolute energy scales, such as particle masses, nuclear and atomic binding energies and energy levels, the neutron lifetime, etc. The most natural way to parametrize the energy scale is through the temperature of the thermal bath, $T_\gamma$. In this case, as we will see shortly, it is more meaningful to parametrize the baryon and dark matter densities by the following parameters, which are proportional to the time-independent baryon-to-photon and dark matter-to-photon \emph{number ratios}:
\be
\varpi_i \equiv \frac{\omega_i}{T_0^3},  \ \ \ \ i = b, c.
\ee
We also define $\varpi_m \equiv \varpi_b + \varpi_{cdm}$. As we will see, all cosmological quantities \emph{at a given $T_\gamma$} depend on the parameters $\varpi_i, \Lambda$, $N_\mathrm{eff}$, etc., but not \emph{directly} on $T_0$. This parametrization of the evolution of the Universe also requires redefining ``comoving" scales as physical scales at a fixed \emph{energy scale}, rather than a fixed time. \\

Of course, cosmological observables depend {\it both} on absolute energy scales and on the expansion history relative to today. Since the two scalings described above are incompatible with each other, a variation of $T_0$ cannot be fully absorbed. Thus the present photon temperature is indeed measurable with cosmological data, independently of FIRAS.

\subsection{Background evolution and last scattering}

We now show that, when expressing background quantities in terms of the photon temperature (rather than time $t$, redshift $z$ or scale factor $a$), following the second scaling described in the previous section, they do not depend on $T_0$, at fixed baryon-to-photon and CDM-to-photon number ratios $\varpi_b, \varpi_{cdm}$. 

First, the Hubble rate $H(T_\gamma)$ is given by 
\be
H^2(T_\gamma) = \frac{\L}{3} + \gamma_{100}^2 \left(\mathcal{A}(T_\gamma)~ T_\gamma^4 + \varpi_m T_\gamma^3 \right),
\label{eq:background}
\ee
where the function $\mathcal{A}(T_\gamma)$ is equal to a fixed number $\bar{\cal A}$ proportional to $1+\frac{7}{8} \left( \frac{4}{11}\right)^{4/3} N_\mathrm{eff}$ until the heaviest neutrino becomes non-relativistic, and later on only depends on the ratio\footnote{or more realistically on $\sum_i m_{\nu i}/T_\gamma$, where the sum runs over all neutrinos that are non-relativistic at a given time.} $m_\nu/T_\gamma$. Thus matter-radiation equality occurs at a temperature $T_{\rm eq}=\varpi_m/\bar{\cal A}$ that only depends on $\varpi_m$. Of course, the \emph{current} Hubble rate $H_0$ does depend on the current CMB temperature $T_0$. Thus one can think of $T_0$ as parametrizing the current age of the Universe. \\

Second, in our baseline $\Lambda$CDM model, primordial nucleosynthesis starts with deuterium fusion, at a temperature fixed by the baryon-to-photon number ratio and the deuterium binding energy. The final abundance of all primordial nuclei depends on nuclear rates and on the neutron-to-proton ratio, governed by the neutron lifetime, the effective number of degrees of freedom of the Standard Model, and the Fermi constant. Thus, in the minimal cosmological model, the only cosmological parameter relevant for this process is the baryon-to-photon number ratio, or equivalently $\varpi_b$. It sets in particular the primordial Helium fraction, $Y_{\rm He}$, which is relevant for cosmological recombination\footnote{The \texttt{CLASS} code~\cite{Blas:2011rf}, that we will use in the following, includes a nucleosynthesis fitting function that adjusts $Y_{\rm He}$ as a function of $\omega_{\rm b}$, assuming the FIRAS value for $T_0$. For the purpose of this work, we modified this fitting function, which now takes $\varpi_b$ as its input parameter.}.\\

\begin{figure*}[ht]
    \centering
    \includegraphics[width = 2 \columnwidth]{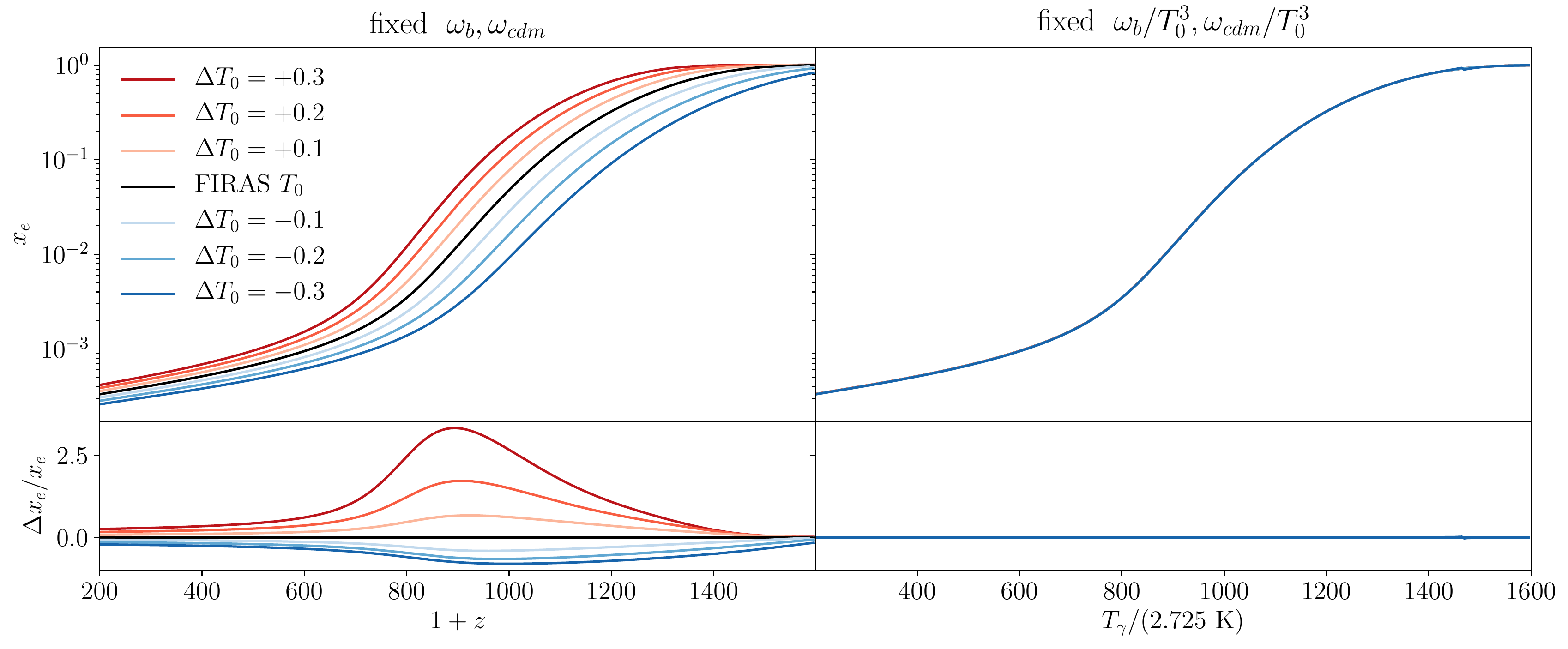}
    \caption{Free-electron fraction $x_e$ for different values of the present-day CMB monopole $T_0$. On the left panel, we keep $\omega_b, \omega_{cdm}$ constant, and show $x_e$ as a function of redshift, which leads to large variations when changing $T_0$, as found in Ref.~\cite{Chluba:2007zz}. On the right panel, we keep $\omega_b/T_0^3, \omega_{cdm}/T_0^3$ constant, and show $x_e$ as a function of photon temperature. In terms of these rescaled variables, $x_e$ is independent of $T_0$. In both cases, the Helium mass fraction $Y_{\rm He}$ is kept constant; this is consistent with BBN predictions only if $\omega_b/T_0^3$ is kept constant.}
    \label{fig:xe}
\end{figure*}

Lastly, cosmological recombination can formally be described by coupled equations of the form (see, e.g.~\cite{2011PhRvD..83d3513A})
\beqa{recomb}
\frac{d x_e}{dt} &=& \mathcal{F}(x_e, T_e, f_\nu, \rho_b, H, T_\gamma),\\
\frac{dT_e}{dt} &=& \mathcal{G}(x_e, T_e, H, T_\gamma), \\
\frac{d f_\nu}{dt} &=& \mathcal{C}(x_e, f_\nu, \rho_b, H, T_\gamma),
\eeqa
where $x_e$ is the free electron fraction, $T_e$ is the electron temperature, $f_\nu$ is the photon phase-space density in the neighborhood of the Lyman-$\alpha$ transition, and the last equation is a Boltzmann equation describing its evolution. Upon rewriting these equations in terms of $dx_e/dT_\gamma, dT_e/dT_\gamma, d f_\nu/d T_\gamma$, and using the fact that $\rho_b \propto \varpi_b T_\gamma^3$ and $H$ is a function of $\varpi_m$ and $T_\gamma$ only at the recombination epoch, we find that the free-electron fraction is a function of $T_\gamma$ and $\varpi_b, \varpi_{cdm}$ only:
\be
x_e = x_e(T_\gamma; \varpi_b,\varpi_{cdm}).
\ee
We show the free-electron fraction for different values of $T_0$ in Fig.~\ref{fig:xe}, where we illustrate that, when keeping constant baryon-to-photon and dark matter-to-photon number ratios and expressing $x_e$ as a function of $T_\gamma$, it is indeed independent of $T_0$.

Given the recombination history $x_e(T_\gamma)$, one can compute the visibility function $g(\eta)$, which is the differential probability of last scattering per unit conformal time $\eta$:
\be
g(\eta) \equiv \dot{\tau} \exp\left[- \int_{\eta}^{\eta_0} d\eta'\dot{\tau}(\eta')\right]
\ee
where $\dot{\tau} = a n_{\rm H} x_e \sigma_{\rm T}$ is the differential Thomson optical depth, and $\eta_0$ is the current conformal time. It peaks at $\eta_*$ such that $g'(\eta_*) = 0$, which thus solves the equation
\be
\frac{d \dot{\tau}}{d \eta}\Big{|}_{\eta_*} = \dot{\tau}^2(\eta_*). \label{eq:eta_s}
\ee
Rewriting $a = T_0/T_\gamma$ and using the fact that $n_\mathrm{H}a^3$ is given by $\omega_b$ (up to a factor predicted by nucleosynthesis), we find that $\dot{\tau}/T_0$ is a function of $T_\gamma, \varpi_b, \varpi_{cdm}$. Using $d/d\eta = a H T_\gamma d/dT_\gamma = T_0 H d/dT_\gamma$, we then find that Eq.~\eqref{eq:eta_s} is satisfied for a temperature $T_\gamma = T_*$ independent of $T_0$, and depending only on $\varpi_b, \varpi_{cdm}$. 
For future reference, 
we write the corresponding fitting function that can be obtained 
by adjusting a numerical fit of Ref.~\cite{Hu:2004kn},
\be 
\label{eq:tstar}
T_* \approx 2970\left(\frac{\varpi_m ~{\rm K}^3}{7.06\times 10^{-3}}\right)^{0.0105}
 \left(\frac{\varpi_b ~{\rm K}^3}{1.1\times 10^{-3}}\right)^{-0.028}\, {\rm K}\,.
\ee
Thus, the \emph{temperature of last-scattering $T_*$ is independent of $T_0$} for some given $\varpi_b, \varpi_{cdm}$. Of course, the \emph{redshift} of last-scattering $z_*$ does depend on $T_0$ through $1 + z_* = T_*/T_0$, and so does the conformal time at last scattering $\eta_* \propto 1/T_0$. 

The effective sound speed of the photon-baryon fluid $c_s = \frac1{\sqrt{3}}\left(1 + \frac34 \frac{\rho_b}{\rho_\gamma}\right)^{-1/2}$ is a function of $T_\gamma, \varpi_b$ only. 
Therefore, the comoving sound horizon at last scattering $r_s$ is such that $T_0 r_s$ is a function of $\varpi_b, \varpi_{cdm}$ only:
\be
r_s = \int_0^{\eta_*} c_s d \eta = \frac1{T_0} \int_{T_*}^{\infty} c_s(T_\gamma;\varpi_b)\frac{dT_\gamma}{H(T_\gamma;\varpi_m)}. 
\ee
This implies that the \emph{physical scale} of the sound horizon at recombination, $a_* r_s = (T_0/T_*) r_s$, is a function of $\varpi_b, \varpi_{cdm}$ only, and does not depend on $T_0$. 

The same argument holds true for the moment of baryon decoupling (also called the baryon drag time), which happens slightly after recombination.
Therefore, the physical size of the
sound horizon at baryon decoupling $r_{d,\,{\rm phys}}$,
which is important for the 
BAO measurements, does not depend on $T_0$.
A useful fitting function 
for $r_{d,\,{\rm phys}}$
can be obtained by 
combining Eq.~\eqref{eq:tstar}
with the fit of
Ref.~\cite{Aubourg:2014yra},
\be 
\label{eq:rd}
\begin{split}
r_{d,\,{\rm phys}} = a_d r_d \approx &~ 0.1386
\left(\frac{\varpi_m ~{\rm K}^3}{7.06\times 10^{-3}}\right)^{-0.26}\\
&\times 
 \left(\frac{\varpi_b ~{\rm K}^3}{1.10\times 10^{-3}}\right)^{-0.10}
~{\rm Mpc}\,. 
\end{split}
\ee 
The CMB anisotropy power spectrum that we will discuss in~\ref{sec:CMB} depends crucially on the physical photon diffusion damping scale at last scattering, which is of the form 
\be 
a_* r_{\rm damp} = 2\pi \left[ 
\int_0^{\eta_*} \frac{{\cal D}(R)}{\dot{\tau}} d\eta \right]^{1/2}~,
\ee
where ${\cal D}$ is a function of $R\equiv\frac{3\rho_b}{4\rho_\gamma}$ (see for instance \cite{Hu:1995em}). Like for $a_* r_s$, some elementary steps show that this physical scale only depends on fundamental constants and on $\varpi_b, \varpi_{cdm}$.

\subsection{Transfer functions and initial conditions} \label{sec:transfer}

Transfer functions are solutions of the cosmological linear perturbation equations for each wavenumber $k$ normalised in the super-Hubble regime, for instance, to ${\cal R}(k)=1$. For fixed initial conditions, CMB anisotropy and large-scale structure observables depend on a number of such transfer functions evaluated at different epochs.

Another potential source of $T_0$-dependence
is the normalization of transfer functions. The cosmological 
perturbations are characterized 
by the conformal comoving wavenumber $k$,
which is equal to the physical wavenumber 
now (which is typically normalized as $a=1$ now). 
Since the current Universe age 
depends on $T_0$, the comoving wavenumbers
depend on it as well, as opposed to the 
\textit{physical} wavenumbers.
However, there is a way to rescale the 
conformal momenta such that the dynamics 
of cosmological perturbations do not depend 
on $T_0$.

It is a straightforward exercise to rewrite equations for linear cosmological perturbations in terms of $T_\gamma$ rather than conformal time (starting from, e.g.~\cite{Ma:1995ey}). By doing this, one can find that the transfer functions depend on $T_\gamma, \varpi_b, \varpi_{cdm}, \L$ and $k/T_0$, where $k$ is the comoving wavenumber. This can be easily understood as follows. Instead of a set of comoving scales $k$ that correspond to inverse physical scales at the current time, one can define another set of comoving scales $\tilde{k}$ coinciding with inverse physical scales \emph{at a fixed photon temperature}. If we choose arbitrarily this temperature to be $T_\gamma = 1$, the two sets are related through $\tilde{k} = k/T_0$.

Then, two universes with the same $\varpi_b, \varpi_{cdm}, \L$ have the same $\tilde{k}$-dependent transfer functions \emph{at a given photon temperature}. Such universes are statistically identical if they further have the same $\tilde{k}$-dependent power spectrum $\Delta^2_{\cal R}$ of scalar fluctuations. In other words, 
the r.m.s.~amplitudes of primordial 
fluctuations should be the same 
on the same physical scales in both universes.
The simple power-law spectrum motivated by inflation is typically defined at some
arbitrary pivot scale $k_{\rm P}$, which 
should be appropriately rescaled in 
a Universe with different $T_0$.
Alternatively, one could rescale the 
amplitude itself. Indeed, keeping $k_{\rm P}$ fixed, we may rewrite the primordial curvature power spectrum as a function of $\tilde{k} = k/T_0$ as follows:
\be 
\Delta^2_{\cal R} = A_s(k/k_{\rm P})^{n_s-1} = A_s T_0^{n_s -1} (\tilde{k}/k_P)^{n_s-1}.
\ee
Thus, in the case of power-law initial conditions, our scaling scheme requires the combination $A_s T_0^{n_s-1}$ to be fixed.

\begin{figure*}[ht]
\begin{center}
\includegraphics[width=0.49\textwidth]{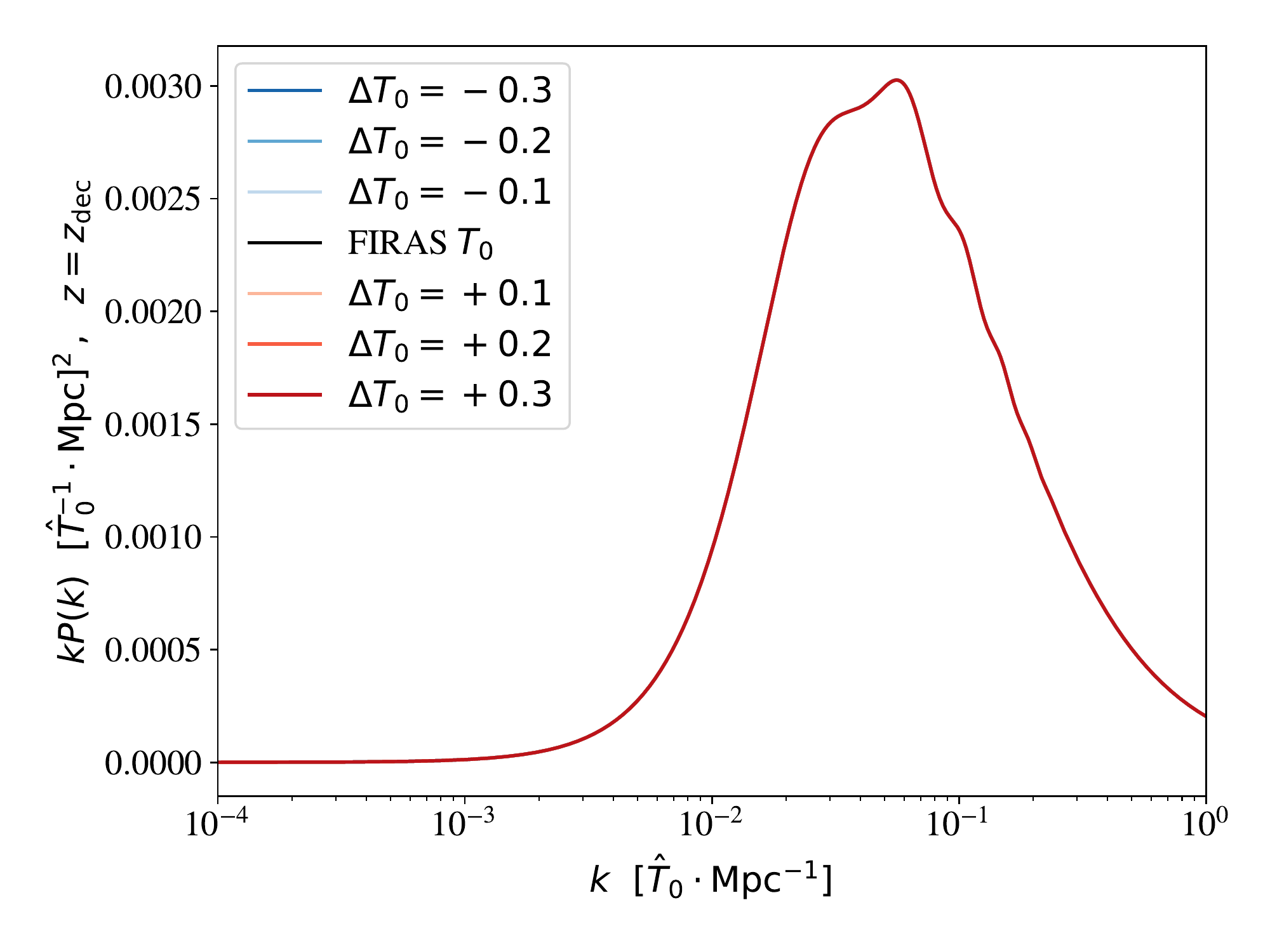}
\includegraphics[width=0.49\textwidth]{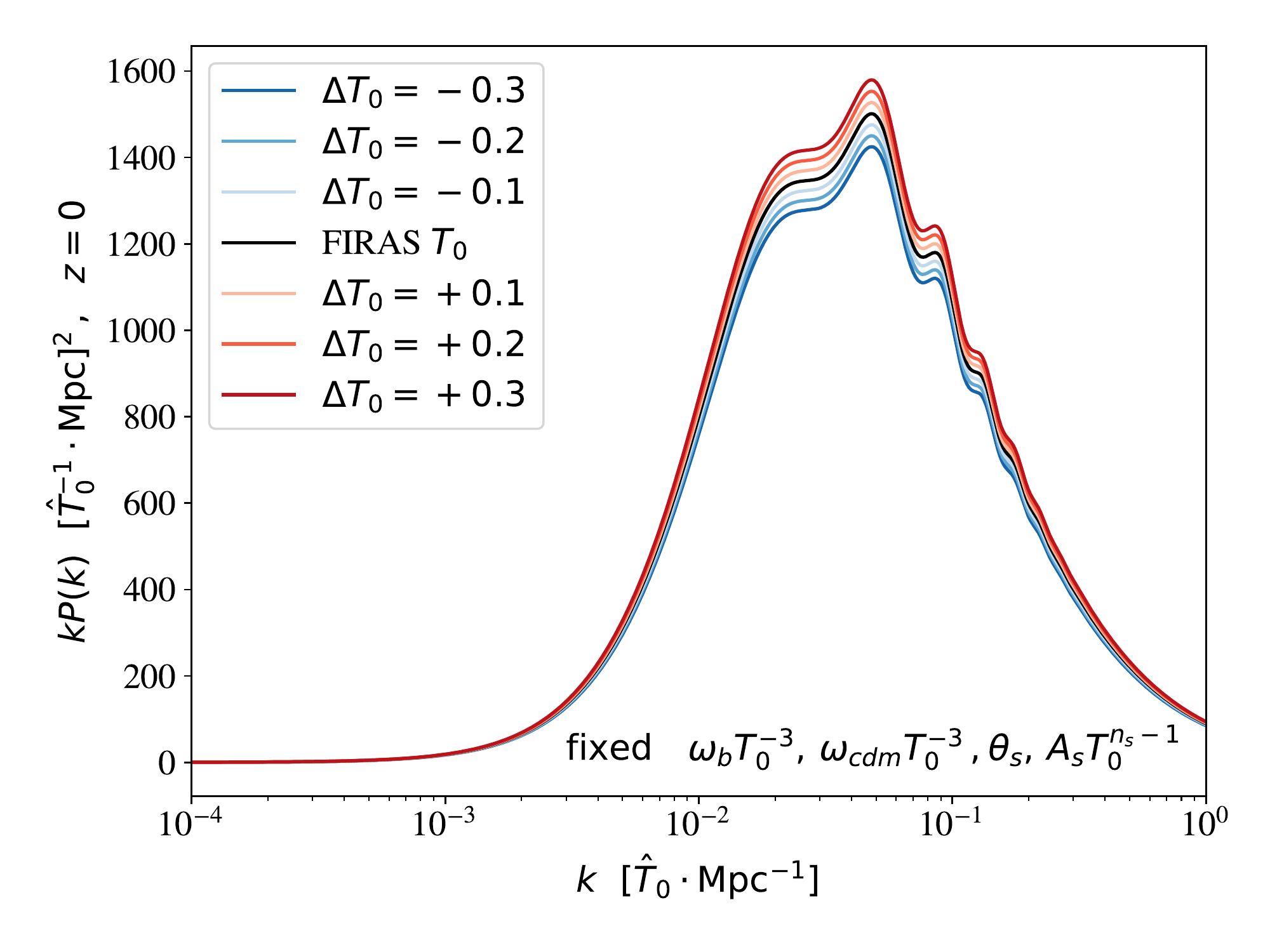}
\end{center}
\caption{ 
Effect of a variation of $T_0$ (quoted in Kelvins) on the matter power spectrum at the time of last scattering (left panel)
and today (right panel). 
The combinations $\omega_b/T_0^3, \omega_{cdm}/T_0^3, \theta_s$ and $A_s T_0^{n_s-1}$ are kept constant. In this case, the matter power spectrum computed at the time of last scattering (left panel)
is independent of $T_0$. 
A constant $\theta_s$ 
fixes the ratio
between $T_0$ and~$\Lambda$.
This choice does not noticeably affect the physics at recombination, but introduces a correlation 
between $T_0$ and the growth 
factor seen when we evaluate the power spectrum
at the present time (right panel). We use the units 
$\hat{T}_0\cdot {\rm Mpc}^{-1}$ ($\hat{T_0} \equiv T_0/T_{0,\,{\rm FIRAS}}$), such that all 
spectra are shown as functions of the same rescaled wavevector $\tilde{k} = k/T_0$. 
}
\label{fig:tovar2}  
\end{figure*}

We illustrate these points in Fig.~\ref{fig:tovar2}, where we show the matter power spectrum for several values of $T_0$, but fixed $\varpi_b, \varpi_c$ and $A_s T_0^{n_s -1}$, as a function of $\tilde{k} = k/T_0$. 
We see that it remains completely unaffected by $T_0$ when computed at the decoupling redshift $z_{\rm dec}$, corresponding to the fixed energy scale $T_*$. However, when computing it \emph{at the present time}, i.e.~at an energy scale $T_0$, its amplitude clearly varies with $T_0$, as expected.

The value of the cosmological constant used to produce these plots is extracted from $\theta_s$, which is fixed to the Planck best-fit value. 
This way, varying $T_0$ changes the amount of time elapsed since recombination until today, which changes the relative current fraction of the cosmological constant energy density w.r.t matter density. This leads to 
a different growth history and hence affects the
amplitude of the power spectrum at $z=0$. 
These effects will be discussed in detail momentarily.

\subsection{Primary CMB anisotropies\label{sec:CMB}}

There are two different contributions to the CMB primary (unlensed) power spectra. On small scales, anisotropies are mostly sourced by photon, baryon and metric fluctuations at last scattering, whose transfer functions contain damped oscillatory features associated to the physical sound horizon scale $a_* r_s$ and damping scale $a_* r_{\rm damp}$.
On large scales, the temperature spectrum receives an additional contribution from the late-time integrated Sachs-Wolfe (ISW) effect (see e.g.~\cite{Gorbunov:2011zzc}), resulting from the time variation of gravitational potentials when the cosmological constant becomes important. 

\begin{figure*}[ht]
\begin{center}
\includegraphics[width = 2\columnwidth]{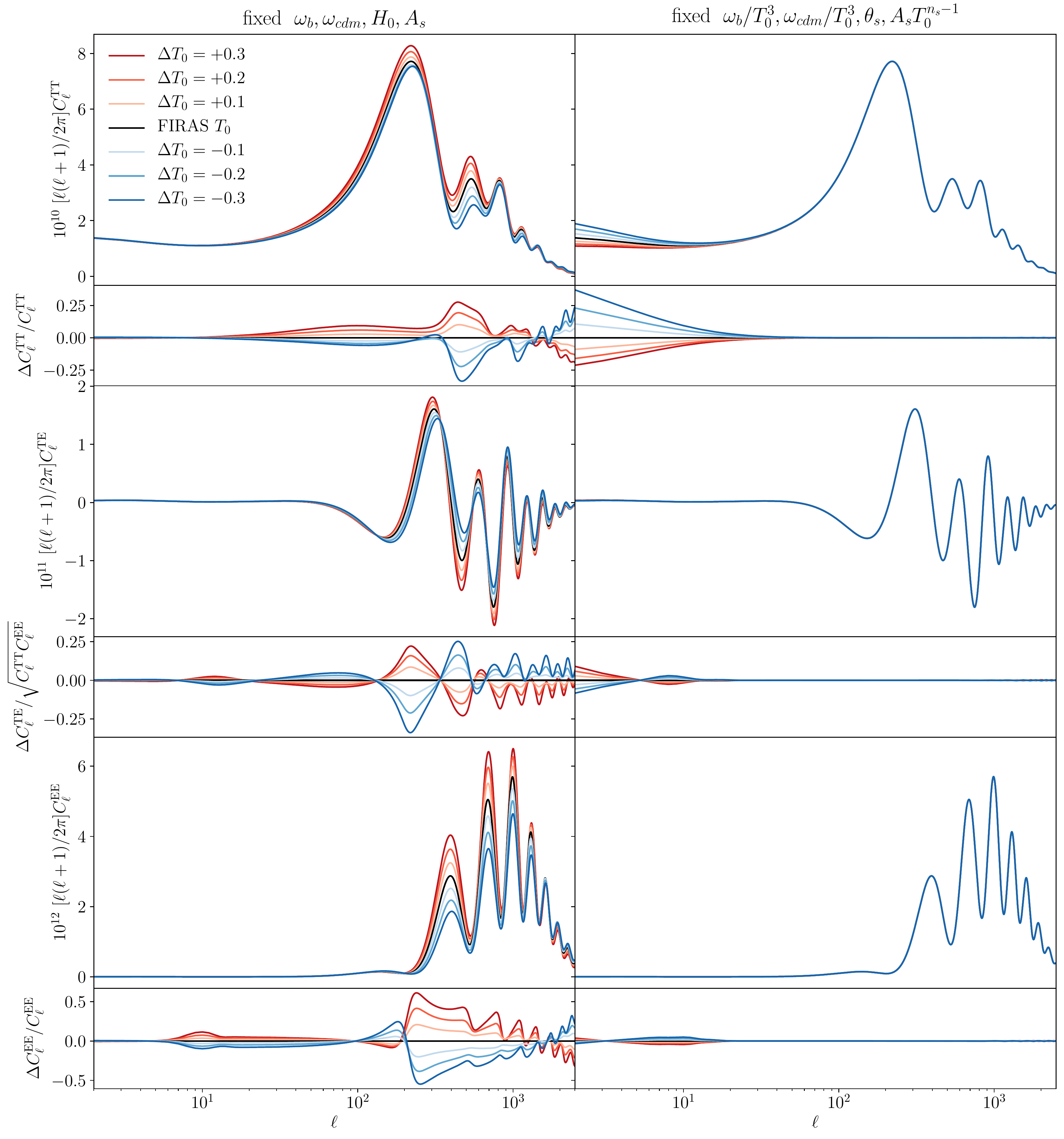}
\end{center}
\caption{Effect of a variation of $T_0$ (quoted in Kelvins) on the lensed TT, TE and EE CMB power spectra. In the left column, $\omega_b, \omega_{cdm}, H_0$ and $A_s$ are kept constant, as in Refs.~\cite{Chluba:2007zz, Hamann:2007sk}. In the right column, the combinations $\omega_b/T_0^3, \omega_{cdm}/T_0^3, \theta_s$ and $A_s T_0^{n_s-1}$ are kept constant. With the latter choice of constant parameters, CMB power spectra are independent of $T_0$ on small scales, but do depend on $T_0$ at large scales through the ISW effect. In both cases, $\tau_{\rm reio}$ is kept constant.
}
\label{fig:tovar} 
\end{figure*}

We have seen that for fixed $\varpi_b$, $\varpi_{cdm}$, $\Lambda$ and $A_s T_0^{n_s-1}$, not only the background evolution $H(T_\gamma)$ is independent of $T_0$, but so are the temperature of last scattering $T_*$, the transfer functions expressed in terms of $(T_\gamma, \tilde{k})$, and all r.m.s. fluctuations as a function of the same two variables. Nevertheless, the unlensed CMB spectra still depend on $T_0$, for the two following reasons:
\begin{enumerate}
\item The CMB spectra are inferred from spherical maps, and thus expanded in multipoles rather than wavenumbers. The (Legendre) transformation from wavenumber to multipole space involves implicitly the angular diameter distance $D_A$, which contains an integral over the expansion history relative to the current time, and thus depends on $T_0$, as anticipated in \ref{sec:is_measurable}. For instance,  the angular diameter distance to the last scattering surface reads 
\be
D_A(T_*) = a_* \int_{T_0}^{T_*} \frac{d T_\gamma}{T_0 H(T_\gamma)} = \frac1{T_*} \int_{T_0}^{T_*} \frac{d T_\gamma}{H(T_\gamma)},
\label{eq:DA}
\ee
and depends explicitly on $T_0$ through the lower integration boundary. This means that the scaling that we discussed so far preserves the exact shape of the CMB unlensed spectra $C_\ell$, on all scales for the polarization spectrum, and on small scales that are unaffected by the late ISW effect for the temperature spectrum. However it shifts these spectra horizontally to larger or smaller multipoles depending on the value of $T_0$. 
\item The late ISW effect depends essentially on the amount of expansion during $\Lambda$ domination, given by $1+z_\Lambda = T_\gamma^\Lambda/T_0$. For a fixed $\Lambda$, the photon temperature at matter-to-$\Lambda$ equality $T_\gamma^\Lambda$ is fixed, but $z_\Lambda$ depends explicitly on $T_0$. Thus late ISW contribution to the large scale temperature spectrum depends on $T_0$.
\end{enumerate}
We show the variations of CMB power spectra in Fig.~\ref{fig:tovar} with different $T_0$. On the left column, we keep $\omega_b, \omega_{cdm}, H_0$ and $A_s$ constant, recovering the results of Refs.~\cite{Chluba:2007zz, Hamann:2007sk}. On the right, we keep $\omega_b/T_0^3, \omega_{cdm}/T_0^3$ and $A_s T_0^{n_s -1}$ constant; we moreover keep the angular scale $\theta_s \equiv a_* r_s/D_A(T_*)$ fixed. The \texttt{CLASS} code~\cite{Blas:2011rf} automatically adjusts $\Lambda$ -- or equivalently, $H_0 \simeq \left[\frac{\Lambda}{3}+\gamma_{100}^2\varpi_m T_0^3\right]^{1/2}$ -- to produce any requested input $\theta_s$. We see that when these parameters are kept fixed, the TT, TE, and EE CMB spectra computed by the \texttt{CLASS} code~\cite{Blas:2011rf} are exactly invariant on small scales when $T_0$ varies. However, this transformation 
preserves neither ($\Lambda$, $T_\gamma^\Lambda$) and the absolute energy scale of $\Lambda$ domination, nor 
($\Omega_\Lambda$, $z_\Lambda$) and the amount of expansion during $\Lambda$ domination. Thus the amplitude of the late ISW effect is different, which can be seen as residual variations on large angular scales in Fig.~\ref{fig:tovar}. 
These large scales are poorly constrained by observations due to the effect of cosmic variance. As a consequence, the large geometric degeneracy between $T_0$ and $H_0$ prevents CMB observations alone from providing tight bounds on either parameter.

We can understand the direction of the $T_0-H_0$ degeneracy analytically as follows.
 The integral in eq.~(\ref{eq:DA}) is dominated by the low-temperature end, for which we may neglect the radiation contribution to $H(T_\gamma)$, i.e.
\be
\label{eq:h}
H(T_\gamma) \approx \gamma_{100} \sqrt{h^2 +  \varpi_m(T_\gamma^3 - T_0^3)}.
\ee
Since $T_* \gg T_0$, we may further take the upper boundary to infinity, and arrive at
\beqa{DA}
&&\gamma_{100} D_A(T_*) \approx \frac1{T_*} \int_{T_0}^{\infty} \frac{d T}{\sqrt{h^2 + \varpi_m(T^3 - T_0^3)}}\nonumber\\
&&= \frac{T_0}{h T*} \int_1^\infty \frac{dx}{\sqrt{1 + \Omega_m(x^3 -1)}} \equiv \frac{T_0}{h T*} \mathcal{I}(\Omega_m),
\eeqa
where $\Omega_m \equiv \varpi_m T_0^3/h^2 = \omega_m / h^2$ is the usual matter density fraction, and the function $\mathcal{I}(\Omega_m)$ can be written in terms of a hypergeometric function.

\begin{figure*}[ht]
\begin{center}
\includegraphics[width=0.6\textwidth]{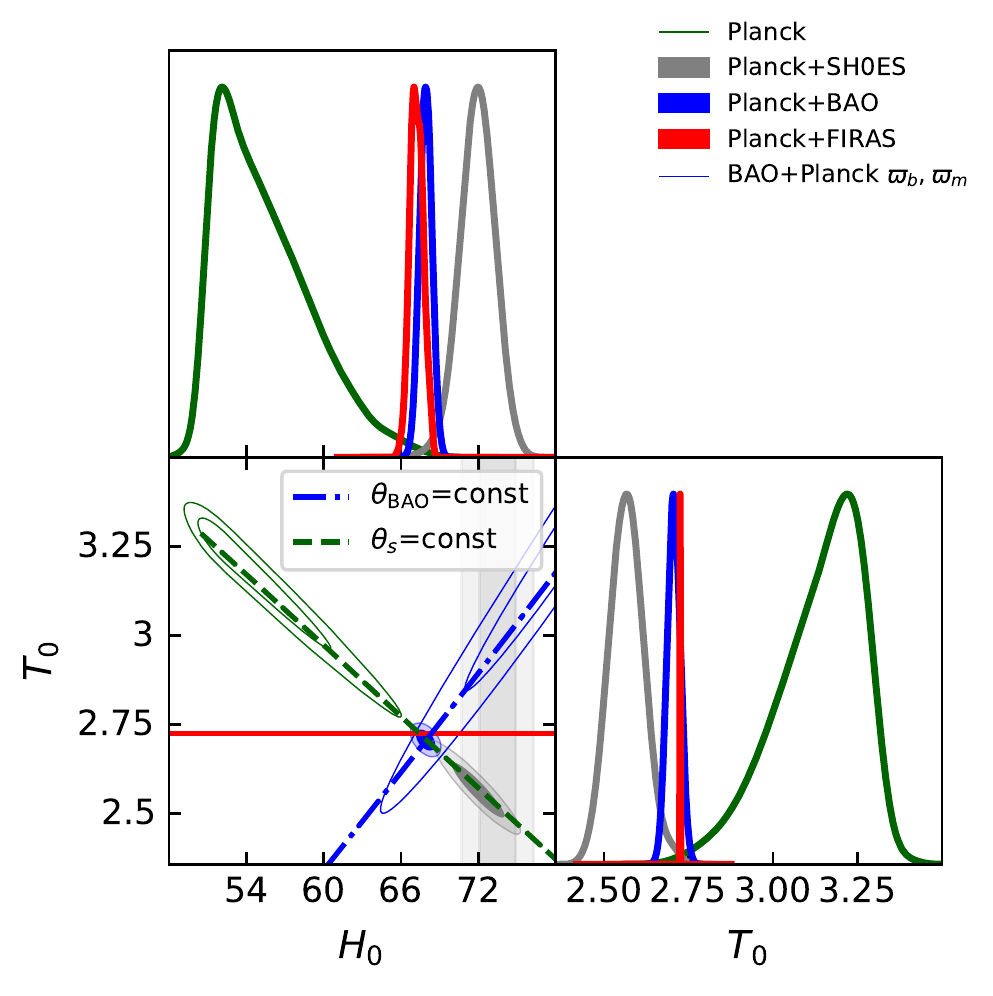}
\end{center}
\caption{
Posterior distribution for $T_0$ and $H_0$
extracted from the analysis of 
the following data sets:
Planck 2018 data,
Planck plus the SH0ES $H_0$
prior, Planck plus
the BOSS DR12 BAO, Planck plus the FIRAS $T_0$
prior, and finally 
the BAO plus the Planck priors on $\varpi_b,\varpi_m$.
We stress that
$T_0$
was varied in all these analyses.
A horizontal solid red line 
marking the FIRAS value is put for illustrative purposes.
The green dashed
line shows 
the degeneracy direction
$\theta_s(H_0,T_0)=$~const (which can be approximated as
$H_0\propto T_0^{-1.2}$ in the vicinity of the Planck+FIRAS best-fitting point),
whereas the blue long dashed line shows the degeneracy $H_0\propto T_0^{0.86}$ expected to produce a fixed BAO acoustic angle $\theta_{\rm BAO}$ (see the main text for
more details). 
\label{fig:small} } 
\end{figure*}

Assuming the Planck+FIRAS best-fit values for $\varpi_m$, $T_0$ and $h$ for numerical calculations, we obtain, 
\be 
\label{eq:dadegen}
\frac{\partial\ln D_A}{\partial\ln T_0}\Bigg|_{\rm Planck} \approx -0.22\,,\quad \frac{\partial\ln D_A}{\partial\ln h}\Bigg|_{\rm Planck} \approx-0.19\,,
\ee
at constant $\varpi_m$. Thus, 
around the Planck best-fit cosmology,
the angular diameter distance is mostly a function of the combination $H_0 T_0^{1.2}$, and we expect an approximate degeneracy $H_0 \propto T_0^{-1.2}$ at fixed $\varpi_b, \varpi_{cdm}$. 
This estimates agrees 
with the degeneracy 
found in our MCMC analysis for small 
variations of $T_0$ and $H_0$ around the
baseline Planck cosmology. To capture larger deviations, we can use the 
exact implicit 
relation
\be 
\theta_s(H_0,T_0)
\Big|_{{\rm fixed}~\varpi_b,\,\varpi_m}=\theta_{s,{\rm best-fit~Planck}}\,,
\ee 
with $\theta_s$ being computed numerically
in \texttt{CLASS}.  
This prediction for the degeneracy direction
agrees well with 
the shape of our MCMC contours 
for the full range of $T_0$ and $H_0$ probed in the analysis, see Fig.~\ref{fig:small}. 
The details of this analysis will be discussed in the next Section.

\subsection{CMB lensing\label{sec:CMBlensing}}

\begin{figure}[ht]
    \centering
    \includegraphics[width=\columnwidth]{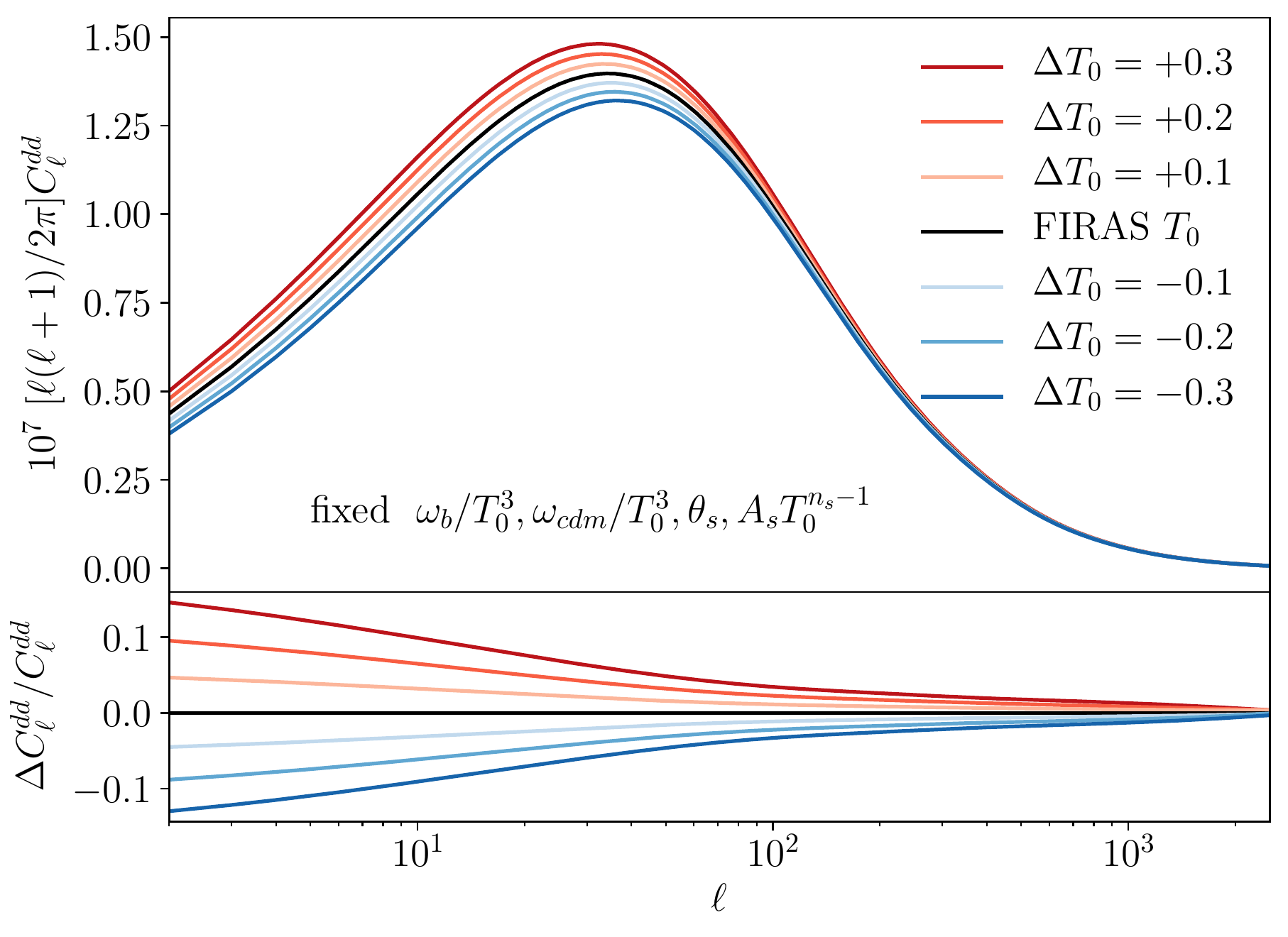}
    \caption{Power spectrum of the lensing deflection angle as a function of $T_0$, for fixed $\omega_b/T_0^3, \omega_c/T_0^3, \theta_s$ and $A_s T_0^{n_s -1}$. Note that this power spectrum is computed entirely within the Limber approximation for simplicity.} \label{fig:lensing}
\end{figure}

We have already seen that keeping $\omega_b/T_0^3$, $\omega_m/T_0^3$ and $\theta_s$ constant does not preserve the amount of expansion taking place between last scattering and matter-to-$\Lambda$ equality, given by $T_*/T_\gamma^\Lambda$, nor during $\Lambda$ domination, given by $T_\gamma^\Lambda/T_0$ (where $T_\gamma^\Lambda$ depends on $T_0$). A different ratio $T_\gamma^\Lambda/T_0$ implies a different amplitude of the late ISW contribution to large-scale temperature anisotropies. The variation of $T_0$ should be further imprinted 
through the CMB lensing effect, which also correlates with the late-time decay factor of metric fluctuations during $\Lambda$ domination. Besides, the CMB lensing spectrum should be shifted horizontally by a different angular diameter distance to small redshifts. As is usually the case when changing the redshift of matter-to-$\Lambda$ equality, these different effects nearly compensate each other at the level of the lensing potential on small angular scales, and appear mainly at $\ell \lesssim 100$ \cite{Lewis:2006fu}. This is confirmed in Fig.~\ref{fig:lensing}, where we show the power spectrum of deflection angle for different values of $T_0$.

The smoothing of acoustic peaks in the CMB temperature and polarization spectra is sensitive to a broad range of multipoles around the peak of the deflection-angle power spectrum $C_l^{dd}$ shown in Fig.~\ref{fig:lensing}. Thus, this smoothing is slightly impacted by a change of $T_0$ when $\omega_b/T_0^3, \omega_{cdm}/T_0^3, \theta_s$ and $A_s T_0^{n_s-1}$ are kept constant. Together with the late ISW effect, this is one of the two mechanisms through which the (lensed) CMB spectra are sensitive to $T_0$. Finally, CMB lensing extraction allows to measure the deflection spectrum and can marginally increase the sensitivity of CMB experiments to $T_0$, although this technique has more sensitivity to scales $\ell \geq 70$ at which the effect of the CMB temperature on $C_l^{dd}$  is gradually suppressed.

In conclusion, we see that the geometric degeneracy of $T_0$ and $H_0$ in primary CMB anisotropies is broken by the ISW effect on large scales, as well as gravitational lensing.
Actually, it gets broken even more clearly when including information on large scale structure observables, as we shall now discuss.

\subsection{Large-scale structure}

Let us discuss now the impact
of varying $T_0$
on large-scale structure.
For simplicity, let us focus on
the baryon acoustic oscillation
(BAO) measurements, which are widely
used to break the geometric 
degeneracy of CMB data.

The galaxy distribution 
mapped by spectroscopic surveys is a three-dimensional observable, which is characterized by angles and redshifts.
This can be contrasted with the CMB,
whose BAO pattern is only observed in 
projection
onto the two-dimensional 
last-scattering sphere. 

The galactic BAO is usually 
probed through the 
position-space two-point correlation
function, which peaks at the spatial
separation corresponding to the 
physical size of the BAO scale at the observed epoch.
This scale is not 
directly observed, because a galaxy
survey only measures the angular 
position of the galaxies and their redshift.
To convert these coordinates into 
a grid of comoving distances one 
typically assumes some fiducial 
cosmology.
As long as the difference between the 
distances in fiducial and true 
cosmologies 
are small, the geodesic distance between a pair 
of galaxies can be described by
the so-called Alcock-Paczynski (AP) scaling 
parameters\footnote{See the original paper~\cite{Alcock:1979mp} 
and Refs.~\cite{Matsubara:1996nf,Ballinger:1996cd} 
for the first applications of the scaling parameters in the form used nowadays, e.g. in the official BOSS data analysis~\cite{Alam:2016hwk}.
},
\be 
\alpha_\parallel = \frac{H_{\rm fid}(z)}{H(z)}\,, \quad 
\alpha_\perp =\frac{D_A(z)} {D_{A,\,\rm fid}(z)}\,.
\ee 
$\alpha_\parallel$ and $\alpha_\perp$
capture, correspondingly, the radial and angular fractions of the 
separation between two galaxies.
The description
in terms of the AP parameters 
is adequate 
for space-times with small spatial curvature gradients 
that behave globally like the Friedman-Robertson-Walker-Lemaitre Universe
\cite{Xu:2012fw,Heinesen:2018hnh,Heinesen:2019phg}. 
 Thus, it will be sufficient 
for our analysis within the flat $\Lambda$CDM
model.

The isotropic component of the galaxy distribution is mostly sensitive to the following combination of the AP parameters:
\be 
\alpha = (\alpha_\parallel \alpha_\perp^2)^{1/3}\,,
\ee 
which describes how a small spatial volume ``reconstructed" from 
the observed volume of angles and redshifts rescales due
to a difference between the true and fiducial
cosmologies.

It is customary to parametrize
the galactic BAO 
with an effective angular size
of the acoustic peak
in the two-point 
correlation function~\cite{Gorbunov:2011zzc},
\be 
\theta_{\rm BAO}=
\frac{r_{d}}{D_V(z_{\rm eff})}\,,
\ee 
where $z_{\rm eff}$ is the 
effective (weighted)
redshift (see e.g. Eq.~(9) from Ref.~\cite{Coil:2003cf})
and $D_V$ is the effective comoving volume-averaged distance to the galaxy sample \cite{Xu:2012fw},
\be 
D_V(z_{\rm eff})
=((1+z_{\rm eff})^2D^2_A(z_{\rm eff})z_{\rm eff}/H(z_{\rm eff}))^{1/3}\,.
\ee 
Physically, $\theta_{\rm BAO}$
can be thought of as a scale 
of the BAO in the 
angular two-point galaxy 
correlation function, 
which is averaged over the redshift bin of a survey. 
It is a very close counterpart of $\theta_s$ measured in the CMB data.
We stress that the AP conversion is only a technical tool to extract this angle from the data.\footnote{Mathematically, it is equivalent to extracting the BAO scale from the angular 
power spectrum of the observed galaxies, as it is done for the CMB. 
An advantage of the distance conversion approach is that it allows one to directly compare the 
data to a 3d power spectrum
model without having to project 
it onto the sky.
A discussion on this point can be found, e.g. in Ref.~\cite{1978ApJ...221....1D}.}

The anisotropic part 
of the galaxy BAO signal is characterized by another combination of the AP 
parameters,
\be 
\epsilon = \left(
\frac{\alpha_\parallel}{\alpha_\perp}
\right)^{1/3}-1\,,
\ee 
which describes how the difference between the true and fiducal cosmology affects the relative scaling between the 
radial and transverse 
distances. 
The combination of 
isotropic and anisotropic BAO signals 
allows one
to separately constrain the parameters 
\be 
\theta_{\rm BAO,\,\parallel}\equiv r_d H(z_{\rm eff})\quad 
\text{and}\quad 
\theta_{\rm BAO,\,\perp}\equiv \frac{r_d}{(1+z_{\rm eff})D_A(z_{\rm eff})} \,.
\ee 
Note that the anisotropic BAO signal 
is a quite weak probe of 
cosmological parameters in minimal models
\cite{Alcock:1979mp,Ballinger:1996cd,Ivanov:2019pdj}.
The base $\Lambda$CDM model with varied
CMB temperature $T_0$
considered in this paper
also belongs to this class.

Importantly, 
the effective redshift 
of a galaxy sample 
$z_{\rm eff}$ is always known because it
is measured directly from the data. 
This should be contrasted 
with the CMB observations.
If the physical sound horizon at decoupling $r_{d,\,{\rm phys}}$ is fixed, 
its comoving size 
depends on the unknown decoupling redshift $z_{\rm dec}$ 
and hence $T_0$,
\be 
\label{eq:rd2}
r_d 
=r_{d,\,{\rm phys}}\Big|_{z_{\rm dec}}(1+z_{\rm dec})\,.
\ee 
The effective volume-averaged distance to the galaxies 
has a different dependence
on cosmological parameters
compared to $D_A(z_{*})$. Indeed, performing a calculation similar to Eq.~\eqref{eq:dadegen} 
we find\footnote{Here we use $z_{\rm eff}=0.38$, the effective redshift of the low-z BOSS galaxy sample \cite{Alam:2016hwk}.}
\be 
\frac{\d \ln D_V}{\d \ln H_0}\Big|_{z_{\rm eff}=0.38}
=-0.78\,,\quad 
\frac{\d \ln D_V}{\d \ln T_0}
\Big|_{z_{\rm eff}=0.38}
=-0.33\,.
\ee 
Combining it with 
Eq.~\eqref{eq:rd}
we obtain 
\be 
\frac{\d \ln \theta_{\rm BAO}}{\d \ln H_0}\Big|_{z_{\rm eff}=0.38}
=0.78\,,\quad 
\frac{\d \ln \theta_{\rm BAO}}{\d \ln T_0}
\Big|_{z_{\rm eff}=0.38}
=-0.67\,.
\ee 
Thus, the BAO angle $\theta_{\rm BAO}$ constrains the
combination $H_0T_0^{-0.86}$,
which
is quite orthogonal to the 
line of constant $H_0T_0^{1.2}$
probed by the CMB.
This allows one to break the degeneracy between $T_0$ and $H_0$
when the galaxy BAO is combined with Planck.

This effect is illustrated in Fig.~\ref{fig:small}, 
where we show the $H_0-T_0$
posterior
extracted from the BOSS DR12
BAO data \cite{Ata:2017dya}.
To obtain this posterior, we have fitted the BAO data with \emph{minimal priors} from Planck 2018 data, namely the baryon- and matter-to-photon ratios $\varpi_b,\varpi_m$.
The use of these priors is motivated by the following argument.
When we eventually combine BAO and Planck 2018,
the $\varpi_b,\varpi_m$ 
limits will be totally 
dominated by Planck,
because they are measured to $1\%$ precision 
from the shape of the CMB spectra independently 
of the late-time geometric expansion.

In passing, it is worth noting that the anisotropic BAO signal, 
in principle, allows one
to separately measure $T_0$ and $H_0$ from the BAO data alone
if the priors on $\varpi_b,\varpi_m$ are imposed.  
Indeed, the full BAO signal 
is summarized in terms of two parameters, 
$r_d/D_{A}(z)$ and $r_d H(z)$, 
which depend on $\varpi_b,\varpi_m, H_0$
and $T_0$ in our model (see
Eq.~\eqref{eq:h} and 
Eq.~\eqref{eq:rd}). Once $\varpi_b,\varpi_m$
are fixed by the priors, we are left with two 
parameters to constrain $H_0$ and $T_0$,
which allows one to eventually 
break the degeneracy between them. 
This explains why the $H_0-T_0$ posterior contour from the BAO in Fig.~\ref{fig:small} 
is not an infinite line. 
However, as we can see from this plot, even though 
the $T_0$ and $H_0$ measurements from the BAO alone are possible in principle,
the resulting constraints are quite loose because the anisotropic part of the BAO signal is a very weak function of cosmological parameters.
In combination with Planck only the best 
measured isotropic BAO part matters.

Additionally, one can constrain the peculiar 
velocity fluctuation r.m.s. $f\sigma_8$
from redshift-space distortions.
All these pieces of information break the geometric degeneracies between the late-time 
parameters $\omega_{cdm},\omega_b,h$ and~$T_0$. 

Alternatively, one can break the geometric degeneracy and measure $T_0$ from the Planck data by the 
local measurement from SH0ES
that yields a direct 
prior on $H_0$.
In this paper we will focus
on these two possibilities: the BAO from galaxy surveys and local measurement of $H_0$ by 
Cepheid-calibrated supernovae.

\subsection{Dependence on the calibration}

Just like any bolometer, the Planck satellite measures a power output $P$, proportional to the electromagnetic intensity, hence temperature, as a function of direction $\hat{n}$. Formally (and dropping additional complexities related to the instrumental beam), we have
\be
P(\hat{n}) = G \times T (t, \hat{n}) = G T_0 \left( 1 +  \frac{\Delta T(\hat{n})}{T_0}\right),
\ee
where $G$ is the instrumental gain. Therefore to extract the temperature fluctuation $\Delta T(\hat{n})/T_0$, the first step is to calibrate the instrument, i.e.~estimate $G T_0$. The lower frequency channels of the Planck HFI instrument (100, 143, 217, and 353 GHz), which are the most relevant for cosmological-parameter estimation, are calibrated by using the time-varying orbital dipole \cite{Adam:2015vua}. The basic idea is as follows: the orbital velocity $\vec{v}_{\rm orb}(t)$ of the satellite leads to a time-varying CMB dipole in the satellite's rest frame, with temperature 
\be
T_{\rm orb}(t, \hat{n}) = \hat{n} \cdot \vec{v}_{\rm orb}(t)~ T_0,  
\ee
leading to a time-varying power output $P_{\rm orb}(t, \hat{n}) = G T_0 (\hat{n} \cdot \vec{v}_{\rm orb}(t))$. The orbital velocity of the satellite is very well known, and as a consequence, it is possible to determine the product $G T_0$ to high accuracy. Therefore, without any prior information on $T_0$, Planck can accurately measure the \emph{relative} (dimensionless) temperature fluctuation $\Delta T(\hat{n})/T_0$. 

In practice, the data is provided in terms of a dimensionfull temperature power spectrum $D_{\ell}^{\rm data}$ (with units of $\mu$K$^2$), under the assumption that $T_0 = T_{0,\rm FIRAS}$, which amounts to mutiplying the relative fluctuations by $T_{0,\rm FIRAS}$. To account for this rescaling, we therefore need to multiply the dimensionless
theoretical $C_\ell$'s computed by \texttt{CLASS} by $T_{0\,,{\rm FIRAS}}^2$:
\be 
\label{eq:planckCal}
D^{\text{\rm theory}}_\ell  \equiv 
T_{0\,,{\rm FIRAS}}^2 \frac{\ell (\ell +1)}{2 \pi} C_\ell^{\rm theory}\,. 
\ee
In the first version of this paper, we had incorrectly multiplied $C_{\ell}^{\rm{theory}}$ by the floating $T_0^2$ rather than $T_{0\,,{\rm FIRAS}}^2$. This would have been appropriate had Planck been calibrated on a source whose absolute brightness is known (such as a planet). This is indeed how the higher frequency channels are calibrated \cite{Adam:2015vua}, but those are mostly relevant to foreground separation, and the scaling \eqref{eq:planckCal} is the one most adequate for cosmological parameter estimation. We explain in Appendix \ref{app:calibration} why this apparently innocuous error led to artificially tighter constraints on $T_0$ from the Planck data alone.

\begin{figure*}[ht]
\begin{center}
\includegraphics[width=1.0\textwidth]{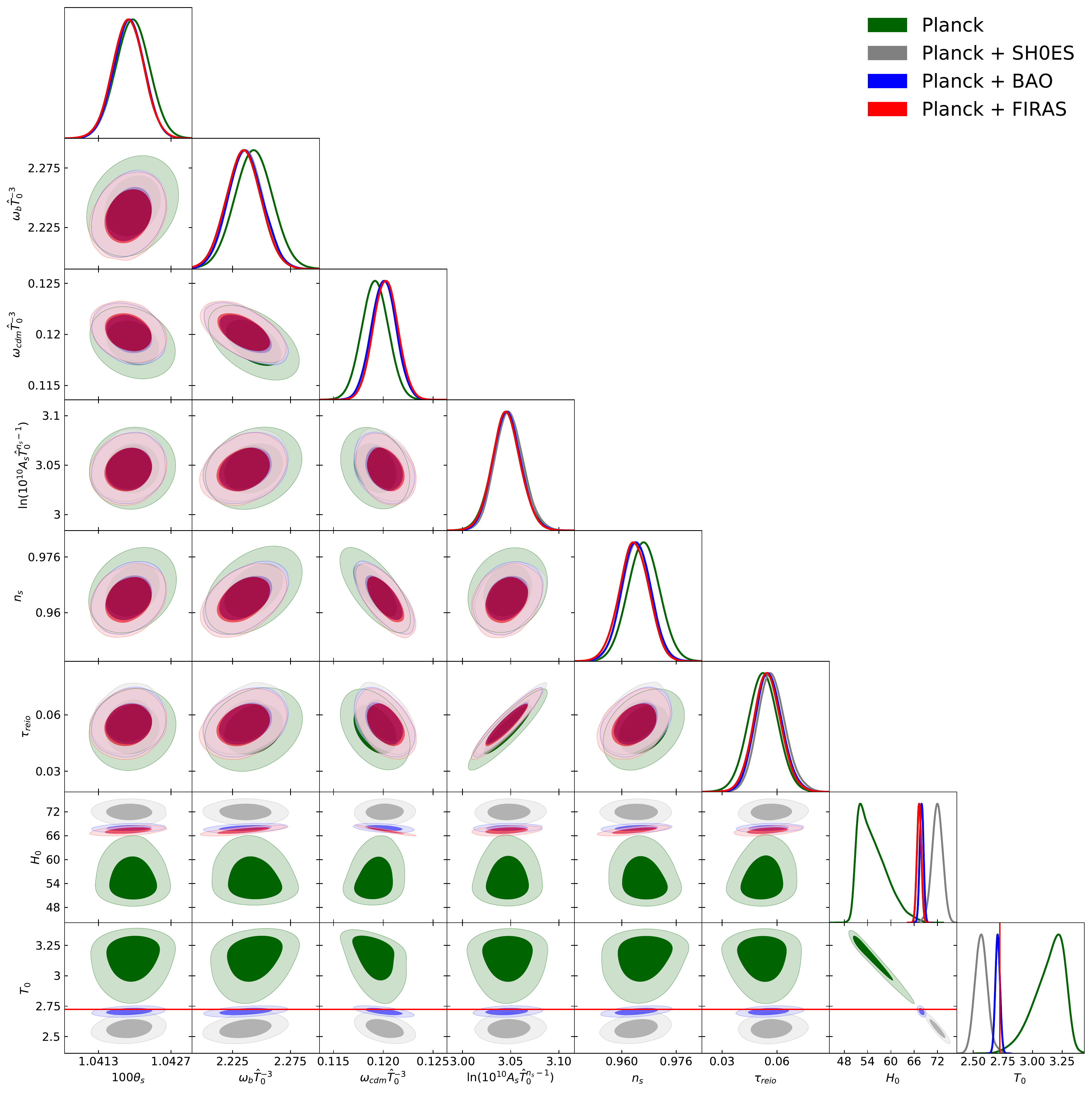}
\end{center}
\caption{Posterior distributions and marginalized 68\% and 95\% confidence contours for the cosmological parameters 
of the $T_0-\L$CDM model fitted to Planck (in green), Planck+SH0ES (in gray), and Planck + BAO (in blue).
For comparison, we also show the result of the Planck baseline analysis with $T_0$ 
fixed to the FIRAS best-fit
value $2.7255$~K (in red).\label{fig:res} } 
\end{figure*}
\begin{table*}[t!]
  \begin{tabular}{|c||c|c|c|c|} \hline
   \diagbox{Parameter}{Dataset}  &  Planck + FIRAS & {\small Planck }  &  {\small Planck~+~SH0ES }   &  
    {\small Planck~+~BAO}  
      \\ [0.2cm]
 \hline 
   $100\,\omega_b
\hat{T}_0^{-3}$   
& $2.235_{-0.014}^{+0.015}$
& $2.243_{-0.015}^{+0.015}$
& $2.236_{-0.015}^{+0.015}$ 
& $2.236_{-0.014}^{+0.015}$ \\ 
\hline
$\omega_{cdm}\hat{T}_0^{-3}$   
& $0.1202_{-0.0012}^{+0.0012}$ 
& $0.1192_{-0.0013}^{+0.0013}$
& $0.1202^{+0.0012}_{-0.0012}$ 
& $0.1200_{-0.0012}^{+0.0012}$ \\ 
\hline 
  $100~\theta_{s}$   & $1.0419_{-0.0003}^{+0.00029}$
  &$1.042_{-0.0003}^{+0.0003}$ 
  & $1.042_{-0.0003}^{+0.00029}$
  & $1.042_{-0.0003}^{+0.0003}$
  \\ \hline
$\tau_{\rm reio}$   & $0.05468_{-0.0078}^{+0.0069}$ 
& $0.05218_{-0.0076}^{+0.0075}$
& $0.05695_{-0.0081}^{+0.007}$
& $0.05546_{-0.0079}^{+0.0072}$
\\ \hline
$\ln(10^{10}A_s \hat{T}^{n_s-1}_0)$   & $3.045_{-0.015}^{+0.014}$ 
& $3.047_{-0.015}^{+0.015}$ 
& $3.046_{-0.015}^{+0.015}$
& $3.046_{-0.015}^{+0.014}$
\\ 
\hline
$n_s$  
&  $0.9637_{-0.0043}^{+0.0041}$
& $0.9664_{-0.0043}^{+0.0043}$
& $0.9641_{-0.0044}^{+0.0043}$
& $0.9642_{-0.0043}^{+0.0043}$
\\ 
\hline
$T_0$   & $2.72548_{-0.00057}^{+0.00057}$ & $3.144_{-0.065}^{+0.17}$& 
$2.564_{-0.051}^{+0.049}$ 
& $2.706_{-0.02}^{+0.019}$ \\   \hline 
\hline
 $100\,\omega_{b}$  &  $2.235_{-0.014}^{+0.015}$ & 
 $3.458_{-0.25}^{+0.54}$
  & $1.865_{-0.12}^{+0.11}$
  & $2.189_{-0.057}^{+0.053}$\\ \hline
     $\omega_{cdm}$  & $0.1202_{-0.0012}^{+0.0012}$ 
  &$0.1836_{-0.011}^{+0.028}$
  & $0.1002_{-0.0058}^{+0.0053}$ 
  & $0.1175_{-0.0021}^{+0.002}$
   \\ \hline
$\Omega_\Lambda$   & $0.6834_{-0.0075}^{+0.0075}$
& $0.27_{-0.27}^{+0.086}$
& $0.7691_{-0.019}^{+0.023}$
& $0.6963_{-0.0088}^{+0.0097}$ \\ \hline
$H_0$   & $67.28_{-0.55}^{+0.53}$ 
& $55.32_{-4.9}^{+1.7}$ 
&$72.01_{-1.3}^{+1.3}$
& $67.92_{-0.51}^{+0.49}$ \\ \hline
$\sigma_8$   & $0.8117_{-0.006}^{+0.0057}$
&$0.6499_{-0.063}^{+0.026}$ 
& $0.8758_{-0.022}^{+0.022}$ 
& $0.8186_{-0.011}^{+0.011}$ \\ 
\hline
\end{tabular}
\caption{Mean values and 68\% CL minimum credible
intervals for the parameters of the $T_0-\L$CDM model fitted to
Planck+FIRAS,
Planck only,
Planck~+~SH0ES, and Planck~+~BAO.
For comparison, we also quote in the leftmost column the results obtained for the baseline Planck analysis with $T_0$ fixed to the FIRAS value 
in the leftmost column.
We assumed flat priors on the first seven parameters, which are defined in the text. The last five rows show derived parameters.
$H_0$ is quoted in km/s/Mpc, $T_0$ is quoted in units of Kelvin, and $\hat{T}_0 \equiv{T_0}/{T_{0,\,{\rm FIRAS}}}$.}
\label{table0}
\end{table*}

\section{Results}
\label{sec:res}

We will now fit our $T_0-\L$CDM model to cosmological data, with a flat prior on the seven parameters motivated by the previous discussion: the CMB temperature itself, $T_0$, plus the six combinations that determine the CMB spectra independently of $T_0$, up to the late ISW and CMB lensing effects:
$\{  \omega_b\hat{T}_0^{-3},\,
\omega_{cdm}\hat{T}_0^{-3},\,
\theta_s,\ln(10^{10}A_s\hat{T}_0^{n_s-1}),\,
\tau_{\rm reio},\,
n_s\}$, with 
$\hat{T}_0=T_0/T_{0,\,{\rm FIRAS}}$.

Note that in section \ref{sec:transfer}, we argued that the quantity fixing the overall normalization of the CMB spectra independently of $T_0$ was $A_sT_0^{n_s-1}$. 
This explains why we take $\ln(10^{10}A_s\hat{T}_0^{n_s-1})$ as one of our basis parameters.

The triangle plots with posterior densities and marginalized 
distributions for the
parameters of our $T_0-\Lambda$CDM model (including the derived parameter $H_0$) are shown in Fig.~\ref{fig:res}. 
For comparison, we also display the contours obtained with a baseline Planck analysis with 
$T_0$ fixed to the FIRAS prior. 
The results of this analysis are in good agreement with the ones 
reported by the Planck collaboration~\cite{Aghanim:2018eyx}.\footnote{We have found some small shifts in the cosmological parameters, which resulted from using \texttt{HyRec} instead of \texttt{recfast} 
\cite{Shaw:2011ez,Ade:2015xua}. These shifts are very small relative to statistical uncertainties.}
~The marginalized limits are presented in Table~\ref{table0}. 

Let us first focus on the Planck-only results. 
The first relevant observation is that 
the posterior distribution of the parameters $\{  \omega_b\hat{T}_0^{-3},\,\omega_{cdm}\hat{T}_0^{-3},\,\ln(10^{10}A_s\hat{T}_0^{n_s-1}), \theta_s\}$
are almost the same in the fit with free $T_0$ and with $T_0$ fixed to the FIRAS value.
There are some small shifts (well below statistical uncertainties) which
are produced by small residual correlations between primary and secondary effects.

We can see that releasing $T_0$ in the fit 
introduces a strong degeneracy direction $T_0-H_0$,
in line with our theoretical arguments. The
resulting posterior for $T_0$ is very non-Gaussian
and peaks at $T_0\approx  3.3$~K, which corresponds to the region of
the parameter space with $\Omega_\Lambda \approx 0$.
It is worth mentioning that we have explicitly imposed a physical 
prior $\Omega_\Lambda \geq  0$ in our MCMC chains, translating\footnote{To get this, we rewrite the condition $\Omega_\Lambda\equiv 1-\Omega_m\geq 0$ as
\be
\begin{split}
1\geq \omega_m h^{-2}\quad 
\Rightarrow 
\quad 
\frac{T_0}{T_{0,~{\rm FIRAS}}}\leq  (h^2 (\omega_m \hat{T}_0^{-3})^{-1})^{1/3}
\end{split}
\ee
and use the best-fit values for $h\approx 0.5$
and $\omega_m \hat{T}_0^{-3}=0.14$.} to 
a prior $T_0\lesssim 3.3$~K.
The skewed shape of the $T_0$ posterior reflects this prior.
The preference for $\Omega_\Lambda=0$ can be 
traced back to the anomalies of the 
Planck data, i.e. the low-$\ell$ deficit and
the lensing anomaly.
Remarkably, releasing $T_0$ 
allows one to fit both these anomalies simultaneously. 
Indeed, increasing $T_0$ suppresses 
the large-scale ISW contribution,
which is preferred by the low-$\ell$ data. Moreover, 
it enhances
the late-time matter clustering and the CMB lensing effect, which is preferred by the observed smoothing of the CMB peaks in the Planck TT spectra\footnote{Note that the latter effect is not fully captured by 
common mass fluctuation amplitude parameter $\sigma_8$
because the physical length corresponding 
to the comoving scale $R=8$~Mpc/h is different in models with different $T_0$ and $H_0$, and thus 
 $\sigma_8$ 
effectively measures the density fluctuations
smoothed at different scales.
When the Planck+FIRAS and Planck $T_0-\Lambda$CDM best-fit models are compared at the same fixed \textit{energy} scales, the 
power spectrum amplitude 
is larger in the $T_0-\Lambda$CDM model, see
Fig.~\ref{fig:tovar2}.
}
We analyze the impact of these anomalies on the eventual
Planck-only constraints in
Appendix~\ref{sec:break}. 

At face value, the Planck-only posterior
for $T_0$ is in tension with FIRAS, and 
the posterior for $H_0$ is in tension with both
Planck+FIRAS and SH0ES. Indeed, the corresponding 
regions of the parameter space have no overlap
with the $95\%$ CL area of the Planck-only
posterior. 
However, there are three
important observations.
First, the $T_0$ measurement turns out to be very sensitive to 
statistical 
anomalies in the Planck data, which may pull the $T_0$ posterior
away from the position expected 
in a fair sample.
Second, the quantification of tensions can be subtle
given a highly-skewed nature 
of the Planck-only probability 
distribution functions for $H_0$
and $T_0$.
Third, one may expect that at least part of the tension
can also be driven by the Bayesian prior volume effects. Indeed, the
peak of the posterior distribution for 
$T_0$ and $H_0$
maps onto the region $\Omega_\Lambda \simeq 0$,
which hits the physical prior $\Omega_\Lambda \geq 0$ that we imposed for this parameter.
It is instructive
to compare the values of $\chi^2$
at relevant best-fit points.
The difference in $\chi^2$
of the Planck-only likelihood 
between the Planck-only 
and Planck + FIRAS best-fitting points is given by\footnote{The contributions from different likelihoods are: 
$-4.96$ (high-$\ell$ TTTEEE); $-0.26$ (low-$\ell$ EE); 
$-1.72$ (low-$\ell$ TT); 
$+1.0$ (lensing).}
\be 
\begin{split}
\Delta \chi^2_{\rm eff} & = \chi^2_{\rm eff} (\text{Planck}) -
\chi^2_{\rm eff}(\text{Planck + FIRAS})~\\
&  = 2769.7 - 2775.64 = -5.94\,.
\end{split}
\label{eq:deltachi2}
\ee 
This difference is similar to typical 
improvements due to a better fit of the anomalies in the Planck spectra, which can be obtained, e.g. 
by including the spatial curvature
or the unphysical lensing smoothing parameter $A_L$
in the $\Lambda$CDM fit \cite{Ade:2015xua,Aghanim:2018eyx}.
In this paper we adopt the point of view that 
these anomalies are statistical fluctuations~\cite{Ade:2015xua,Aghanim:2018eyx,Addison:2015wyg,Aghanim:2016sns}.
Therefore, in order to reduce their effect one should 
combine the Planck data with other data sets.
Indeed, since $T_0$ manifests itself only though the late-time expansion
effects, late-Universe probes like BAO or SH0ES will be able to efficiently break the $H_0-T_0$ degeneracy and reduce the dependence of the Planck-only constraints
on the internal anomalies. 
The situation here is similar to the inclusion of the BAO data in the analysis with floating curvature parameter $\Omega_k$, which allows one to remove the anomalous preference of the
primary CMB 
data for a non-zero spatial curvature~\cite{Ade:2015xua,Aghanim:2018eyx}.

The geometric degeneracy between $H_0$ and $T_0$ gets broken once we impose the $H_0$ prior
from SH0ES. 
The effective $\chi^2_{\rm eff}$ of the Planck likelihoods computed at the best-fitting point to the Planck+SH0ES dataset is slightly worse than that of the concordance model -- or in other words, than that computed at the best-fitting point to Planck+FIRAS -- by:
\begin{eqnarray}
\Delta \chi^2_{\rm eff} &\equiv&  \chi^2_{\rm eff}(\text{Planck+SH0ES})-\chi^2_{\rm eff}(\text{Planck+FIRAS})\nonumber \\
&=&2778.82-2775.64=3.18\,.
\end{eqnarray}

In the Planck+SH0ES analysis, the best-fit $T_0$ is significantly lower than the FIRAS value. This was to be expected from the CMB geometric degeneracy $H_0 \propto T_0^{-1.2}$, which requires a smaller $T_0$ in order to be consistent with the larger $H_0$ prior from SH0ES, given the very well constrained angle $\theta_s$. The optimal values of
$T_0$  from Planck + SH0ES and FIRAS are separated by $\Delta T_0 = 0.16$~K, which corresponds to $3.3$ standard deviations (adding in quadrature the Planck + SH0ES and the FIRAS errors).
We see that by floating $T_0$, we have traded the usual 4.1$\sigma$ Hubble tension\footnote{Computed from $H_0=67.4\pm0.5$~km/s/Mpc (68\%CL) for Planck in the baseline $\Lambda$CDM model.} between Planck+FIRAS and SH0ES for a 3.2$\sigma$ $T_0$ tension between FIRAS and Planck+SH0ES.
 
If we use the BAO to break the geometric degeneracy instead of SH0ES, we obtain 
a measurement of $T_0$ that 
agrees with the FIRAS measurement within $68\%$ CL. 
The goodness of fit to Planck data does not degrade much in this case,
\begin{eqnarray}
\Delta \chi^2_{\rm eff} &\equiv&  \chi^2_{\rm eff}(\text{Planck+BAO})-\chi^2_{\rm eff}(\text{Planck+FIRAS})\nonumber \\
&=&2776.28-2775.64=0.64\,.
\end{eqnarray}
Overall, we observe good agreement between Planck+BAO and FIRAS.

In this work, for concision, we did not discuss the constraints coming from the measurement of the amplitude of the matter power spectrum near the present time. We can briefly mention that weak lensing surveys often report an estimate of the parameter combination $S_8\equiv\sigma_8 (\Omega_m/0.3)^{0.5}$, and that the degeneracy discussed in this work is such that smaller values of $T_0$ lead to smaller values of $S_8$. For instance, our $T_0-\L$CDM best-fit model to Planck+SH0ES has $S_8 \simeq 0.77$, in very good agreement with KIDs and DES measurements \cite{Drlica-Wagner:2017tkk,Wright:2018nix,Asgari:2019fkq}. Thus this model would not be disfavored on the basis of weak lensing data only -- but as we explained, it is disfavoured instead by BAO and FIRAS data.

\section{Conclusions\label{sec:concl}}

In this paper, we have given a detailed description of the effects of the CMB temperature monopole $T_0$ on cosmological observables, namely CMB anisotropies and large-scale structure. We have shown that cosmological background quantities and perturbations are independent of $T_0$ when computed at fixed baryon-to-photon and dark matter-to-photon \emph{number ratios}, and at fixed energy scales. The \emph{observed} CMB anisotropy still depends on $T_0$ because this parameter quantifies the energy scale at the present time, from which observations are carried. The leading effect of $T_0$ on CMB anisotropies is to change the angular diameter distance to the surface of last scattering, which is degenerate with a change of the Hubble parameter $H_0$. This geometric degeneracy approximately translates to the parameter degeneracy $H_0 \propto T_0^{-1.2}$. 

The standard procedure to break the CMB-anisotropy geometric degeneracy is to include the very tight $T_0$ measurement from FIRAS. This leads to a measurement of the Hubble parameter $H_0$, with the well-known tension with the SH0ES measurement. In this work, we considered whether removing the FIRAS prior on $T_0$ and breaking the geometric degeneracy by different means might help alleviate the Hubble tension.

First, we showed, for the first time, that CMB anisotropy data \emph{alone} can be used to measure $T_0$ and $H_0$ simultaneously.
Indeed, the geometric degeneracy is not exact: it is broken on large angular scales by the integrated Sachs-Wolfe effect, and by gravitational lensing
on small angular scales. 
From Planck data alone, we 
can estimate $T_0$ with $\sim 4\%$ precision,
$T_0 =  3.144_{-0.065}^{+0.17}$~K (68\%CL). Due to the $T_0-H_0$ geometric degeneracy, this high optimal value for $T_0$ is paired with a rather low  Hubble constant, $H_0 = 55.3^{+1.7}_{-4.9}$ km/s/Mpc. These values are in clear tension with both FIRAS and SH0ES, but we showed that this is driven by the fact that large $T_0$ / low $H_0$ provides a better fit to internal anomalies in CMB data, namely the ``lensing anomaly'' and the ``low-$\ell$ deficit'', which might just be statistical flukes.
Other fundamental cosmological parameters (such as the baryon-to-photon number ratio) are mostly unaffected. 

Second, rather than breaking the geometric degeneracy by including the FIRAS prior on $T_0$, thus inferring $H_0$, we use the SH0ES prior on $H_0$ to obtain, for the first time, an \emph{independent measurement} of $T_0$ from Planck and SH0ES, $T_0 = 2.56 \pm 0.05$ K. This measurement is in  significant (3.3$\sigma$) tension with the FIRAS measurement. Thus, the $H_0$ tension between (Planck + FIRAS) and SH0ES can be fully recast as a $T_0$ tension between (Planck + SH0ES) and FIRAS. This simple result should serve as a reminder that the fundamental culprits of tensions are not necessarily any single parameter whose measurements differ between different data sets.

One may also break the $T_0-H_0$ geometric degeneracy by including BAO data, as was done in past analyses. In that case, one finds that the resulting $T_0$ is consistent with that measured by FIRAS, and the measured $H_0$ is in tension with SH0ES. The BAO measurement thus seems to arbiter in favor of Planck + FIRAS, and disfavor SH0ES. Still, the Hubble tension -- perhaps better named the Hubble-Penzias-Wilson tension -- remains to be definitively solved.

\vspace{1cm}
\section*{Acknowledgments}

We thank Antony Lewis for pointing out our incorrect handling of $T_0$ in the calibration
of CMB temperature maps, in the first version of this paper.
We also thank 
Jens Chluba, 
Simone Ferraro,
Raphael Flauger,
Daniel Green, 
Luke Hart, 
Jan Hamann,
Colin Hill, 
Carlos Martins,
Alessandro Melchiorri
and 
Yvonne Wong
for their useful 
feedback on the first 
version of our paper.
We are grateful to Matias Zaldarriaga and David Spergel for valuable discussions.
MI is partially supported by the Simons Foundation's Origins of the Universe program. YAH is supported by NSF grant number 1820861. 

Parameter estimates presented in this paper are obtained with the \texttt{CLASS} 
Boltzmann code \cite{Blas:2011rf} interfaced with the \texttt{Montepython} MCMC sampler \cite{Audren:2012wb,Brinckmann:2018cvx}. 
The plots with posterior densities and marginalized limits are generated with the latest version of the \texttt{getdist} package\footnote{\href{https://getdist.readthedocs.io/en/latest/}{
\textcolor{blue}{https://getdist.readthedocs.io/en/latest/}}
}~\cite{Lewis:2019xzd},
which is part of the \texttt{CosmoMC} code \cite{Lewis:2002ah,Lewis:2013hha}. 

\appendix 

\section{Dependence of constraints on the calibration} \label{app:calibration}

In the first version of this paper, we had used $D_{\ell}^{\rm incorrect} = T_0^2 \frac{\ell(\ell+1)}{2\pi} C_{\ell}^{\rm theory}$ in lieu of Eq.~\eqref{eq:planckCal}. In this appendix we explain why this incorrect scaling leads to artificially tighter bounds on $T_0$ from Planck data alone.

We know that in our parameter basis, the overall amplitude of the dimensionless temperature and polarization power spectra $C_{\ell}^{\rm theory}$ depends on the combination $\tilde{A}_s \equiv A_s T_0^{n_s-1}$ (strictly speaking, for $\ell \gtrsim 14$, it depends on the combination $\tilde{A}_s e^{-2 \tau_{\rm reio}}$, but we can consider $\tau_{\rm reio}$ to be approximately fixed by small-$\ell$ polarization data). When we fit the correctly normalized spectrum $D_{\ell}^{\rm correct} = T_{0, \rm FIRAS}^2 \frac{\ell(\ell+1)}{2\pi} C_{\ell}^{\rm theory}$ to the Planck data, we implicitly fix the value of $\tilde{A}_s$. In this case, the $T_0-H_0$ degeneracy is only lifted by the late ISW effect and the CMB lensing effect, as mentioned in sections \ref{sec:CMB}, \ref{sec:CMBlensing}. As can be seen from Figs.~\ref{fig:tovar} and \ref{fig:lensing}, these effects moreover have a rather mild dependence on $T_0$. In particular, the dependence of the lensing deflection spectrum $C_\ell^{dd}$ on $T_0$ quickly drops at scales $\ell \gtrsim 70$.
We have seen that the Planck lensing likelihood probes $C_\ell^{dd}$ mainly at $\ell \geq 70$, and thus, is weakly sensitive to $T_0$.

Instead, when one fits to the data the incorrectly normalized spectrum $D_{\ell}^{\rm incorrect}$, it is the combination $T_0^2 \tilde{A}_s$ that is tightly determined by the overall amplitude of CMB primary spectra. The lensing deflection spectrum $C_\ell^{dd}$ is still proportional to $\tilde{A}_s = (T_0^2 \tilde{A}_s)/T_0^2$. As a consequence, varying $T_0$ at fixed $T_0^2 \tilde{A}_s$ leads to an overall rescaling of the lensing deflection spectrum by $T_0^{-2}$.
This significantly increases the impact of $T_0$ on the smoothing of the acoustic peaks via lensing. It also boosts the sensitivity of the Planck lensing likelihood to $T_0$.
These effects artificially increase the sensitivity of CMB-only data to $T_0$, roughly by a factor 3.

\begin{figure}[h]
    \centering
    \includegraphics[width = \columnwidth]{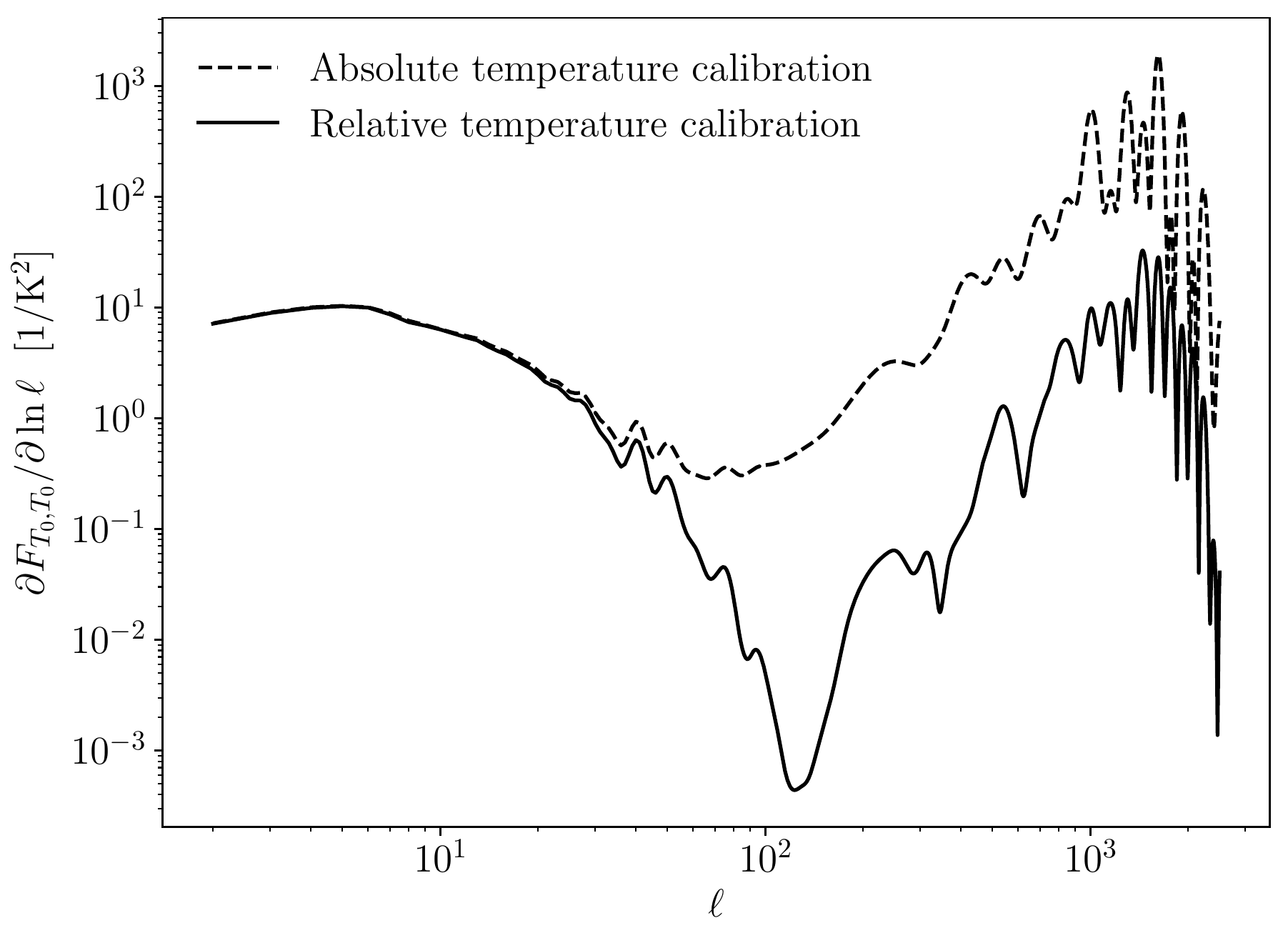}
    \caption{Contribution of the $T_0-T_0$ Planck Fisher matrix element per $\ln \ell$, with the correct relative calibration scaling $D_{\ell} = T_{0, \rm FIRAS}^2 C_{\ell}$ (solid), and when incorrectly using the absolute calibration scaling $D_{\ell} = T_0^2 C_{\ell}$ (dashed). The $\ell \lesssim 10^2$ region corresponds to the ISW effect, whose contribution is mostly unaffected, and the $\ell \gtrsim 10^2$ corresponds to the lensing smoothing effect, artificially enhanced with the incorrect normalisation.}
    \label{fig:Fisher}
\end{figure}

We can make this argument more quantitative with a simple Fisher analysis, accounting for the Planck instrumental noise and angular resolution as given in Ref.~\cite{PlanckBlueBook}. The elements of the 7$\times 7$ Fisher matrix of cosmological parameters take the form $F_{ij} = \sum_{\ell} \partial_{\Omega_i}  C_{\ell}~ \boldsymbol{\mathcal{C}}_{\ell}^{-1} ~\partial_{\Omega_j} C_{\ell}$, where $\boldsymbol{\mathcal{C}}_{\ell}^{-1}$ is the $\ell$-dependent covariance matrix of the lensed TT, TE and EE power spectra (note that for simplicity, we do not quantify here the additional sensitivity to $T_0$ coming from the CMB lensing likelihood, and thus do not consider the lensing deflection spectrum). The variance of parameter $i$ is then given by $(F^{-1})_{ii}$. Note that only the elements of the Fisher matrix are additive in $\ell$, but not the elements of its inverse. We show in Fig.~\ref{fig:Fisher} the contribution of the $T_0-T_0$ element of the Fisher matrix per $\ln \ell$, with the two different calibration scalings. When accounting for the correct scaling, we find that the low-$\ell$ ISW bump and the  high-$\ell$ lensing smoothing have roughly comparable contributions to $F_{T_0 T_0}$. With the incorrect scaling, the contribution of lensing smoothing is significantly boosted and dominates $F_{T_0 T_0}$, resulting in a significant decrease in the error bar on $T_0$. Note that $\textrm{var}(T_0) = (F^{-1})_{T_0T_0} \neq 1/F_{T_0 T_0}$, but the two are close for small correlations.

\section{Quantifying different contributions in the Planck-only constraints}
\label{sec:break}

In this appendix we study the different contributions that
form the Planck-only constraints on $T_0$. Moreover,
we quantify the sensitivity of these constraints to the 
low-$\ell$ deficit 
and the lensing anomalies.
To that end we ran three
additional analyses based on Planck-only data, which we compare with the baseline case called here ``low-$\ell$~+~high-$\ell$~+~lensing'': 
\begin{enumerate}
\item 
    The case ``$\tau$-prior~+~high-$\ell$~+~lensing'' includes all data from Planck except the two low-$\ell$ (TT, EE) likelihoods, which cover $\ell\leq 29$. Since these low-$\ell$ likelihoods are the only ones constraining $\tau_{\rm reio}$, we add to this data set a Gaussian prior $\tau_{\rm reio}=0.0544\pm 0.075$; however any information on the low-$\ell$ deficit and on the late ISW effect is removed.
     \item The case ``low-$\ell$~+~high-$\ell$'' includes all data from Planck except the lensing likelihood derived from 4-point correlation functions. CMB lensing is then only probed through the smoothing of the TT, TE, EE spectra.
    \item The case `` low-$\ell$~+~high-$\ell$~+~$A_L$'' includes the same data (all Planck data except the lensing likelihood) and features one additional and unphysical parameter $A_L$, that rescales the lensing deflection spectrum, and thus parametrizes the amplitude of the smoothing of the acoustic peaks due to CMB lensing.
\end{enumerate}

\begin{figure*}[ht]
\begin{center}
\includegraphics[width=0.6\textwidth]{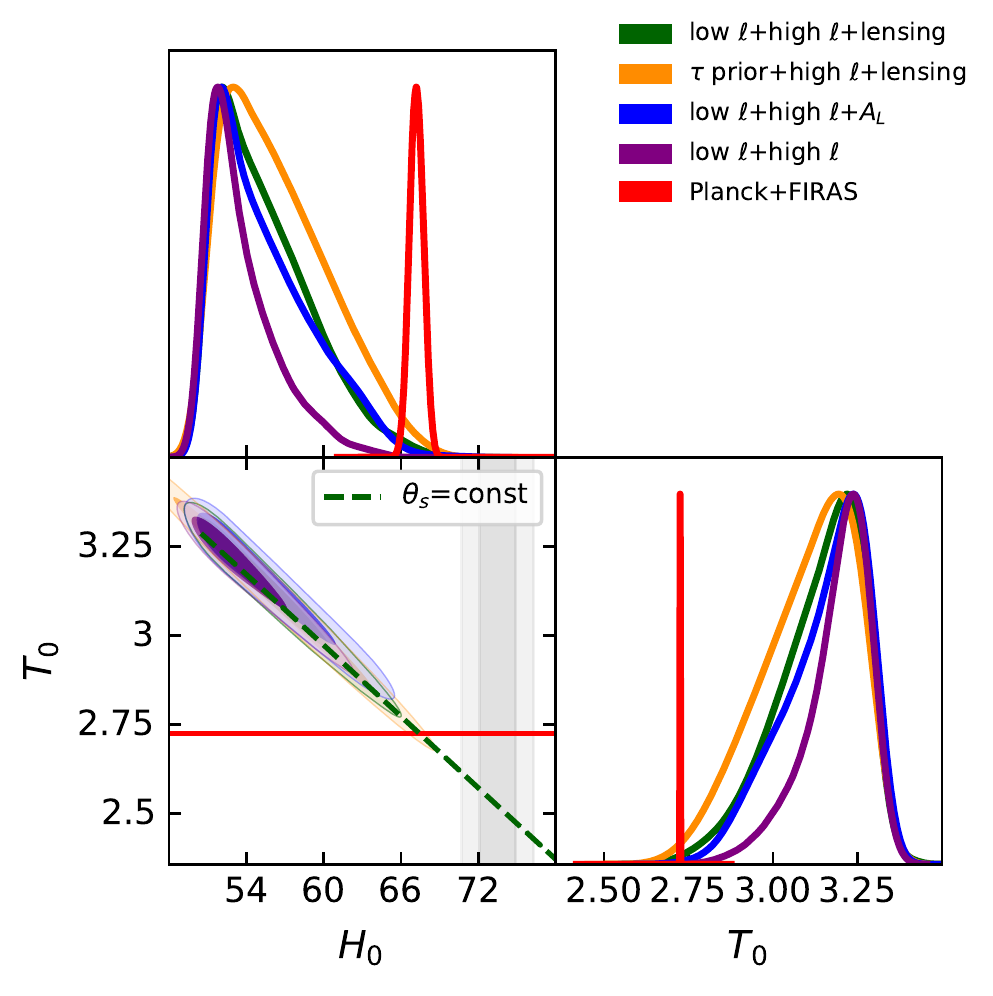}
\end{center}
\caption{
Posterior distribution for $T_0$ and $H_0$
extracted from the analysis of 
the following Planck data sets: low-$\ell$
+high-$\ell$
+lensing, $\tau$-prior+high-$\ell$ [$\ell\geq 30$]+lensing,
low-$\ell$+high-$\ell$,
 low-$\ell$+high-$\ell$+free $A_L$, and full Planck (low-$\ell$
+high-$\ell$
+lensing)+FIRAS,
see the main text for more detail.
A horizontal solid red line 
marking the FIRAS value is put for illustrative purposes.
The green dashed
line shows 
the degeneracy direction
$\theta_s(H_0,T_0)=$~const, the vertical gray band displays the SH0ES $H_0$ prior. 
\label{fig:break} } 
\end{figure*}

The comparison of the first case with the baseline is intended to isolate the impact of the late-time ISW effect.
The comparison of the second case with the baseline isolates the impact of CMB lensing, as probed by the measurement of the lensing deflection spectrum with the 4-point correlation functions. Finally, the comparison of the second and third case is aimed to isolate to some extent the impact of CMB lensing as probed instead by the smoothing of the TT, TE and EE spectra.
The results of our analyses
are summarized in Table~\ref{table1}
and in Fig.~\ref{fig:break}. 

The comparison of the first case with the baseline (orange versus green curves) confirms that the late ISW effect contributes to the preference for a large $T_0$, since the 68\% lower bound on $T_0$ increases by 0.046K (i.e. by about $\sim 0.6\sigma$) when the low-$\ell$ data is restored. We have seen that a large $T_0$ implies a lower ISW component in the temperature spectrum, and thus provides a better fit to the low-$\ell$ data (in particular, to the low quadrupole). 
However, without low-$\ell$ data, there is still a clear preference for a large $T_0=3.110^{+0.20}_{-0.077}$K (68\%CL, orange curve), suggesting that the deficit in the low-$\ell$ data is not the main effect that pushes $T_0$ above the standard value. We get an independent confirmation of this conclusion when comparing the likelihood of the best-fit $\Lambda$CDM and $\Lambda$CDM+$T_0$ models, both fitted to the full Planck baseline data. Out of the total $\Delta \chi^2\approx -6$ reported in eq.~(\ref{eq:deltachi2}), only $\Delta \chi^2 \approx -2$ comes from the two low-$\ell$ likelihoods. This suggests that the late ISW effect and the low-$\ell$ deficit contribute by only $\sim 30\%$ to the preference for a large $T_0$. Since $T_0$ impacts CMB data only through the late ISW and CMB lensing effects, the remaining 70\% can either be driven by {\it (i)} the measurement of the lensing deflection spectrum, or {\it (ii)} the smoothing of the acoustic peak in the TT, TE, EE spectra.

To evaluate the impact of the former, we can compare the second case with the baseline results (purple versus green curves). Actually, when the lensing likelihood is removed, the preference for a high $T_0$ is stronger: the 68\% lower bound on $T_0$ increases by 0.056K (i.e. by about $\sim 0.8\sigma$). This suggests that the lensing spectrum data prefers standard values of $T_0$, and thus, the lensing smoothing must be the main effect responsible for the large $T_0$ preference. Once more, this can be confirmed by the comparison of the likelihood of the best-fit $\Lambda$CDM and $\Lambda$CDM+$T_0$ models, both fitted to the full Planck baseline data. The best fit model of the extended case slightly degrades the lensing likelihood by $\Delta \chi^2 \approx +1$, while it improves the high-$\ell$ likelihood by $\Delta \chi^2 \approx -5$. Looking at the impact of $T_0$ on $C_l^{dd}$ (see figure \ref{fig:lensing}), we conclude that Planck data on the extracted lensing spectrum does not favor the enhancement of the peak caused by large values of $T_0$.

Since large $T_0$ values are driven by the high-$\ell$ likelihood, which is only sensitive to $T_0$ through lensing smoothing effects, there must be a strong connection between  the preference for a large $T_0$ and the well-known ``lensing anomaly'', i.e. that fact that high-$\ell$ Planck data can be better fitted when assuming a larger lensing smoothing effect than that predicted by the concordance $\Lambda$CDM model. It is thus interesting to compare cases two and three, which contain the same amount of Planck data (i.e. everything except the lensing likelihood), but differ through the floating of the $A_L$ parameter. If the effect of $A_L$ and $T_0$ were exactly equivalent, and if $T_0$ was left totally free, we would expect the $\Lambda$CDM+$T_0$+$A_L$ model to be fully compatible with $A_L=1$, and to fit the high-$\ell$ likelihood as well as the $\Lambda$CDM+$A_L$ model.
This is not exactly the case because the prior $\Omega_\Lambda>0$ implies a cut around $T_0<3.3$K that prevents $T_0$ to account for the full effect of $A_L=1.19$ (which is the best-fit value for this data set). The shape of the $T_0$ posterior suggests that without the prior $\Omega_\Lambda>0$, $T_0$ would be driven to even higher values, and would produce even more lensing smoothing. As a result, in the $\Lambda$CDM+$T_0$+$A_L$ fit, $A_L$ remains larger than one (but the physical value $A_L=1$ is now in 1.7$\sigma$ agreement with the data). Another way to see this is that the best-fit $\Lambda$CDM+$T_0$ model fitted to the ``low-$\ell$~+~high-$\ell$'' data reduces the $\chi^2$ of the high-$\ell$ likelihood by $\Delta \chi^2\approx -5$, while with the same data, the best-fit $\Lambda$CDM+$A_L$ model reduces it by $\Delta \chi^2 \approx -7.5$ (section 3.7 of \cite{legacy}). In the $\Lambda$CDM+$T_0$+$A_L$ fit, $T_0$ is more compatible with the FIRAS value than when $A_L$ is fixed to one, but larger values of $T_0$ are still preferred; this comes from a volume effect in the marginalization of the $T_0$ posterior over $A_L$, bearing in mind that in this fit, the role of increasing the lensing smoothing effect is shared between the two parameters $A_L$ and $T_0$.

As a final comment, we note that the $\Lambda$CDM+$T_0$ model (which is physical, although in strong tension with direct $T_0$ measurements) does a better job in fitting the full Planck data than the $\Lambda$CDM+$A_L$ model (which is unphysical). This comes from the fact that $T_0$ increases the lensing deflection spectrum only on large scales (see figure \ref{fig:lensing}), while $A_L$ increases it on all scales. Large values of both parameters are favored because they produce more smoothing in the TT, TE, EE spectrum (the smoothing depends mainly on the amplitude of $C_\ell^{dd}$ around the peak), but large values of $A_L$ are more penalised by the lensing likelihood. Indeed, this likelihood is sensitive to $C_\ell^{dd}$ on smaller angular scales -- at which the effect of $A_L$ is clearly disfavored while $T_0$ has a smaller no impact. Again this can be seen very well by comparing best-fit $\chi^2$'s. When the $\Lambda$CDM+$A_L$ is fitted to the full Planck data, it reduces the best-fit $\chi^2$ by only -3.4 (section 3.39 of \cite{legacy}), because large values of $A_L$ are strongly penalized by the lensing likelihood. Instead, the $\Lambda$CDM+$T_0$ model reduces the best-fit $\chi^2$ by -6 because large values of $T_0$ are less penalized by the lensing likelihood.

Overall, the observed trends suggest the following picture:
the bias of $T_0$ toward high values is driven by the 
Planck high-$\ell$ likelihood
and can be partly understood as a result of the ``lensing anomaly.''
This shift also produces a better fit to the low-$\ell$
data. The only data set that 
pulls $T_0$ (and $H_0$) in the opposite direction toward the Planck+FIRAS concordance model is the CMB lensing reconstruction 
data.

\begin{table*}[t!]
  \begin{tabular}{|c||c|c|c|c|c|} \hline
   \diagbox{ {\small Parameter}}{\small Dataset}  &  Planck + FIRAS & ~{\small low $\ell$+high $\ell$+lens. }  &  ~{\small high $\ell$+lens.+$\tau$}   &  
   ~low $\ell$+high $\ell$
   & 
~ low $\ell$+high $\ell$+$A_L$
      \\ [0.2cm]
 \hline 
   $100\,\omega_b
\hat{T}_0^{-3}$   
& $2.235_{-0.014}^{+0.015}$
& $2.243_{-0.015}^{+0.015}$
& $2.243_{-0.015}^{+0.015}$ 
& $2.242_{-0.015}^{+0.015}$ &$2.256_{-0.017}^{+0.017}$ \\ 
\hline
$\omega_{cdm}\hat{T}_0^{-3}$   
& $0.1202_{-0.0012}^{+0.0012}$ 
& $0.1192_{-0.0013}^{+0.0013}$
& $0.1192^{+0.0013}_{-0.0013}$ 
& $0.1196_{-0.0013}^{+0.0013}$
& $0.1183_{-0.0015}^{+0.0015}$
\\ 
\hline 
  $100~\theta_{s}$   & $1.0419_{-0.0003}^{+0.00029}$
  &$1.042_{-0.0003}^{+0.0003}$ 
  & $1.042_{-0.0003}^{+0.00031}$
  & $1.042_{-0.0003}^{+0.0003}$
  & $1.042_{-0.00031}^{+0.00031}$
  \\ \hline
$\tau_{\rm reio}$   & $0.05468_{-0.0078}^{+0.0069}$ 
& $0.05218_{-0.0076}^{+0.0075}$
& $0.05533_{-0.0071}^{+0.0071}$
& $0.05412_{-0.0076}^{+0.0077}$
& $0.0507_{-0.0079}^{+0.0086}$
\\ \hline
$\ln(10^{10}A_s \hat{T}^{n_s-1}_0)$   & $3.045_{-0.015}^{+0.014}$ 
& $3.047_{-0.015}^{+0.015}$ 
& $3.052_{-0.013}^{+0.013}$
& $3.053_{-0.016}^{+0.016}$
& $3.042_{-0.017}^{+0.017}$
\\ 
\hline
$n_s$  
&  $0.9637_{-0.0043}^{+0.0041}$
& $0.9664_{-0.0043}^{+0.0043}$
& $0.9658_{-0.0044}^{+0.0046}$
& $0.9655_{-0.0044}^{+0.0045}$
& $0.9695_{-0.0051}^{+0.005}$
\\ 
\hline
$T_0$   & $2.72548_{-0.00057}^{+0.00057}$ & $3.144_{-0.065}^{+0.17}$& 
$3.110_{-0.077}^{+0.2}$
& $3.195_{-0.06}^{+0.011}$ &
$3.162_{-0.063}^{+0.16}$
\\   \hline 
$A_L$   & $-$ & $-$& 
$-$
& $-$ &
$1.119_{-0.07}^{+0.063}$
\\   \hline 
\hline
 $100\,\omega_{b}$  &  $2.235_{-0.014}^{+0.015}$ & 
 $3.458_{-0.25}^{+0.54}$
  & $3.355_{-0.29}^{+0.64}$
  & $3.623_{-0.25}^{+0.38}$
  & $3.538_{-0.25}^{+0.54}$\\ \hline
     $\omega_{cdm}$  & $0.1202_{-0.0012}^{+0.0012}$ 
  &$0.1836_{-0.011}^{+0.028}$
  & $0.1782_{-0.013}^{+0.034}$
  & $0.1932_{-0.0081}^{+0.02}$ &$0.1854_{-0.011}^{+0.028}$
   \\ \hline
$\Omega_\Lambda$   & $0.6834_{-0.0075}^{+0.0075}$
& $0.27_{-0.27}^{+0.086}$
& $0.31_{-0.23}^{+0.17}$
& $0.19_{-0.17}^{+0.08}$ & $0.25_{-0.25}^{+0.085}$
\\ \hline
$H_0$   & $67.28_{-0.55}^{+0.53}$ 
& $55.32_{-4.9}^{+1.7}$ 
& $56.29_{-5.8}^{+2}$
& $55.36_{-3.2}^{+1.3}$ &
$55.26_{-4.7}^{+1.7}$
\\ \hline
$\sigma_8$   & $0.8117_{-0.006}^{+0.0057}$
&$0.6499_{-0.063}^{+0.026}$ 
& $0.6645_{-0.075}^{+0.029}$ 
& $0.6325_{-0.042}^{+0.024}$
& $0.6409_{-0.059}^{+0.025}$
\\ 
\hline
\end{tabular}
\caption{Mean values and 68\% CL minimum credible
intervals for the parameters of the $T_0-\L$CDM model fitted to
various data sets:
low-$\ell$
+high-$\ell$
+lensing, $\tau$-prior+high-$\ell$ [$\ell\geq 30$]+lensing,
low-$\ell$+high-$\ell$,
 low-$\ell$+high-$\ell$+free $A_L$, and full Planck (low-$\ell$
+high-$\ell$
+lensing)+FIRAS,
see the main text for more detail.
We assumed flat priors on the first seven parameters, which are defined in the text. The last five rows show derived parameters.
$H_0$ is quoted in km/s/Mpc, $T_0$ is quoted in units of Kelvin, and $\hat{T}_0 \equiv{T_0}/{T_{0,\,{\rm FIRAS}}}$.}
\label{table1}
\end{table*}

\bibliography{short.bib}

\end{document}